\documentclass[showpacs,aps,prd,reprint,superscriptaddress,nofootinbib,longbibliography]{revtex4-1}
\usepackage[colorlinks=true, pdfstartview=FitV,bookmarks=true, bookmarksnumbered=true, breaklinks]{hyperref}
\usepackage[dvipdfmx]{graphicx}
\usepackage{amsmath,amssymb,bm,color,longtable,mathrsfs,slashed,comment,tikz}
\definecolor{phlox}{rgb}{0.87, 0.0, 1.0}
\definecolor{capri}{rgb}{0.0, 0.75, 1.0}
\definecolor{carminered}{rgb}{1.0, 0.0, 0.22}
\hypersetup{linkcolor=carminered, citecolor=phlox, urlcolor=capri}

\def\m#1{\mathrm{#1}}

\definecolor{lime}{HTML}{A6CE39}
\DeclareRobustCommand{\orcidicon}{
	\hspace{-3mm}
	\begin{tikzpicture}
	\draw[lime, fill=lime] (0,0) 
	circle [radius=0.16] 
	node[white] {{\fontfamily{qag}\selectfont \tiny ID}};
	\draw[white, fill=white] (-0.0625,0.095) 
	circle [radius=0.007];
	\end{tikzpicture}
	\hspace{-3mm}
}

\foreach \x in {A, ..., Z}{\expandafter\xdef\csname orcid\x\endcsname{\noexpand\href{https://orcid.org/\csname orcidauthor\x\endcsname}
			{\noexpand\orcidicon}}
}

\begin{document}

\begin{flushright}
DESY~20-235~~~KEK-TH-2286
\end{flushright}

\title{Lattice-fermionic Casimir effect and topological insulators}

\author{Tsutomu~Ishikawa\orcidA{}}
\email[]{tsuto@post.kek.jp}
\affiliation{Graduate University for Advanced Studies (SOKENDAI), Tsukuba, 305-0801, Japan}
\affiliation{KEK Theory Center, Institute of Particle and Nuclear
Studies, High Energy Accelerator Research Organization (KEK), Tsukuba, 305-0801, Japan}

\author{Katsumasa~Nakayama\orcidB{}}
\email[]{katsumasa.nakayama@desy.de}
\affiliation{NIC, DESY Zeuthen, Platanenallee 6, 15738 Zeuthen, Germany}

\author{Kei~Suzuki\orcidC{}}
\email[]{k.suzuki.2010@th.phys.titech.ac.jp}
\affiliation{Advanced Science Research Center, Japan Atomic Energy Agency (JAEA), Tokai 319-1195, Japan}

\date{\today}

\begin{abstract}
The Casimir effect arises from the zero-point energy of particles in momentum space deformed by the existence of two parallel plates.
For degrees of freedom on the lattice, its energy-momentum dispersion is determined so as to keep a periodicity within the Brillouin zone, so that its Casimir effect is modified.
We study the properties of Casimir effect for lattice fermions, such as the naive fermion, Wilson fermion, and overlap fermion based on the M\"obius domain-wall fermion formulation, in the $1+1$-, $2+1$-, and $3+1$-dimensional space-time with the periodic or antiperiodic boundary condition.
An oscillatory behavior of Casimir energy between odd and even lattice size is induced by the contribution of ultraviolet-momentum (doubler) modes, which realizes in the naive fermion, Wilson fermion in a negative mass, and overlap fermions with a large domain-wall height.
Our findings can be experimentally observed in condensed matter systems such as topological insulators and also numerically measured in lattice simulations.
\end{abstract}

\maketitle

\section{Introduction} \label{Sec_1}

The Casimir effect~\cite{Casimir:1948dh,Mostepanenko:1988bs,Bordag:2001qi,Milton:2001yy} is one of the important physical phenomena especially for microscopic systems with spatial boundary conditions.
Although the Casimir effect was originally predicted as early as 1948~\cite{Casimir:1948dh}, the first successful experiment was reported fifty years later~\cite{Lamoreaux:1996wh}.
The original Casimir effect was discussed for the photon field, which is described by quantum electrodynamics (QED), but similar concepts can be extended to any field including scalar, fermion~\cite{Johnson:1975zp,Mamaev:1980jn}, and other gauge fields, which have been actively studied.

Lattice field theories have been broadly used not only as models to study lattice systems realized in solid-state physics but also as tools to simulate more general (quantum) field theories.
A lattice formulation can allow us to investigate physical systems without loss of the nonperturbative effects by using numerical methods such as Monte-Carlo simulations.
Lattice simulations of quantum chromodynamics (QCD), which is the fundamental theory to describe the dynamics of quarks and gluons, are successful examples.

So far, Casimir(-like) effects on the lattice were numerically studied for scalar field theories~\cite{Chernodub:2019kon} and $U(1)$ gauge field theory~\cite{Pavlovsky:2009kg,Pavlovsky:2010zza,Pavlovsky:2011qt} as a simple system.
As nonperturbative field theories, the compact $U(1)$ gauge theory~\cite{Pavlovsky:2009mt,Chernodub:2016owp,Chernodub:2017mhi,Chernodub:2017gwe} and non-Abelian gauge theories such as $SU(2)$~\cite{Chernodub:2018pmt,Chernodub:2018aix} and $SU(3)$~\cite{Kitazawa:2019otp} gauge fields are also studied.
More complicated and interesting examples are systems with an interaction between different fields.
For example, QCD includes a strong coupling between quarks and gluons, which leads to various nonperturbative phenomena such as the confinement, chiral symmetry breaking, and instantons.
Therefore the roles of the Casimir effects in such interacting fermionic systems will be interesting (e.g., see Refs.~\cite{Kim:1987db,Song:1990dm,Song:1993da,Kim:1994es,Vshivtsev:1995xx,Vdovichenko:1998ev,Vshivtsev:1998fg,Braun:2004yk,Braun:2005gy,Braun:2005fj,Abreu:2006pt,Ebert:2008us,Palhares:2009tf,Abreu:2009zz,Abreu:2010zzb,Hayashi:2010ru,Ebert:2010eq,Braun:2010vd,Abreu:2011zzc,Ebert:2011tt,Braun:2011iz,Flachi:2012pf,Flachi:2013bc,Tiburzi:2013vza,Tripolt:2013zfa,Phat:2014asa,Ebert:2015vua,Almasi:2016zqf,Flachi:2017cdo,Nitta:2017uog,Abreu:2017lxf,Wang:2018ovx,Wang:2018qyq,Ishikawa:2018yey,Inagaki:2019kbc,Xu:2019gia,Abreu:2019czp,Ishikawa:2019dcn,Abreu:2019tnf,Abreu:2020uxc,Wan:2020vaj}).

We should mention the other important context of the Casimir effect on the lattice.
The various lattice fermions such as the staggered fermion~\cite{Kogut:1974ag,Susskind:1976jm}, Wilson fermion~\cite{Wilson:1975,Wilson:1977}, and domain-wall (DW) fermion~\cite{Kaplan:1992bt,Shamir:1993zy,Furman:1994ky} also appear in condensed matter physics such as Dirac or Weyl semimetals, topological insulators and ultra-cold atom systems.
For example, the low-energy band structure in Dirac semimetals~\cite{Shuichi_Murakami_2007,Young_2012,Armitage_2018} is Dirac-like (or linear-like), so that it can be regarded as a dispersion relation of relativistic lattice fermion.
Also, the mechanism of gapless surface modes induced from the gapped bulk fermions of topological insulators is formally the same as the chiral fermions realized in the domain-wall fermion formulation.
Our motivation is not limited to theoretical interests and is devoted to future experiments for these condensed-matter materials.
If we can experimentally prepare sufficiently small materials, the Casimir effect for Dirac-like lattice fermions should influence the thermodynamic and transport observables.

In Ref.~\cite{Ishikawa:2020ezm}, we proposed an analytical definition of the Casimir energy for lattice fermions for the first time,\footnote{As early works about the Casimir effect for lattice scalar fields, see Refs.~\cite{Actor:1999nb,Pawellek:2013sda}.} and the various phenomena based on the Casimir energy were investigated.
As a sequential investigation, in this paper, we focus on the following new subjects not studied in Ref.~\cite{Ishikawa:2020ezm}:
\begin{itemize}
\item[(1)]{\it Relationship between Casimir energy and dispersion relation}---We propose that the structure of energy-momentum dispersion relation for a particle gives us a clear and intuitive interpretation of the Casimir effect on the lattice.
In particular, we will emphasize novel phenomena induced by contributions from the doubler modes
\item[(2)]{\it Dependence on spatial dimensions}---We investigate the Casimir effects in the $2+1$- and $3+1$-dimensional space-time while we studied only the $1+1$ dimensions in Ref.~\cite{Ishikawa:2020ezm}.
\item[(3)]{\it Other types of lattice fermions}---For example, Wilson fermions with a negative mass are interesting in the sense that they are closely related to the band structures of topological insulators.
\item[(4)]{\it Derivation from the Abel-Plana formulas}---The Abel-Plana formula is used as a mathematical technique to derive the Casimir effect in the continuum theory.
We give a derivation of the Casimir effect on the lattice by using similar formulas.
\end{itemize}

The theoretical construction of lattice fermions is closely related to the Nielsen-Ninomiya no-go theorem~\cite{Nielsen:1980rz,Nielsen:1981xu}.
This theorem states that when we naively discretize the space, the additional degrees of freedom which are the so-called doubler particles appear~\cite{Wilson:1974sk}.
To eliminate the contribution from doublers in the continuum limit, we can use several fermion formulations.
The Wilson fermion~\cite{Wilson:1975,Wilson:1977}, overlap fermion~\cite{Neuberger:1997fp,Neuberger:1998wv}, and domain-wall fermoin~\cite{Kaplan:1992bt,Shamir:1993zy,Furman:1994ky} are typical examples of the fermion formulation without doublers in the continuum limit.
The Wilson fermion introduces a small but explicit breaking term of the chiral symmetry to evade the no-go theorem.
The Wilson fermion has difficulty in physics which strongly relates to the chiral symmetry but is useful for other physics, and then it is broadly applied for numerical simulations of the QCD.
The overlap fermion is a more sophisticated formulation that can define the chiral symmetry on the lattice.
It also introduces small breaking terms of the chiral symmetry, but we can discuss the details of the chiral symmetry based on the Ginsparg-Wilson relation~\cite{Ginsparg:1981bj}, which represent the chiral symmetry on the lattice.
The domain-wall fermion produces a representation of the overlap fermion.
The M\"obius domain-wall (MDW) fermion~\cite{Brower:2004xi,Brower:2005qw,Brower:2012vk} is an improved formalism of the domain-wall fermion.

This paper is organized as follows.
In Sec.~\ref{Sec_2}, we formulate the Casimir energy of lattice fermions, based on the definition given by Ref.~\cite{Ishikawa:2020ezm}.
In Sec.~\ref{Sec_3}, we see the results for the naive lattice fermions in the $1+1$-, $2+1$-, and $3+1$-dimensional space-time.
In Sec.~\ref{Sec_4}, we investigate the Wilson fermion with a positive or negative mass.
The latter corresponds to the Casimir effect for the bulk modes of topological insulators.
Section~\ref{Sec_5} is devoted to the overlap fermion with the MDW kernel operator, which corresponds to the Casimir effect for surface modes of topological insulators.
In Sec.~\ref{Sec_6}, we summarize our conclusion.
In Appendix~\ref{App:1}, we give a derivation of the Casimir energies for the free massless fermion in the continuum limit with the periodic and antiperiodic boundary conditions.
In Appendices~\ref{App:2}--\ref{App:5}, we give derivations of the Casimir energy for the lattice fermions from the Abel-Plana formulas.

\section{Definition of Casimir energy on the lattice} \label{Sec_2}
\begin{figure}[t!]
    \begin{minipage}[t]{1.0\columnwidth}
        \begin{center}
            \includegraphics[clip, width=1.0\columnwidth]{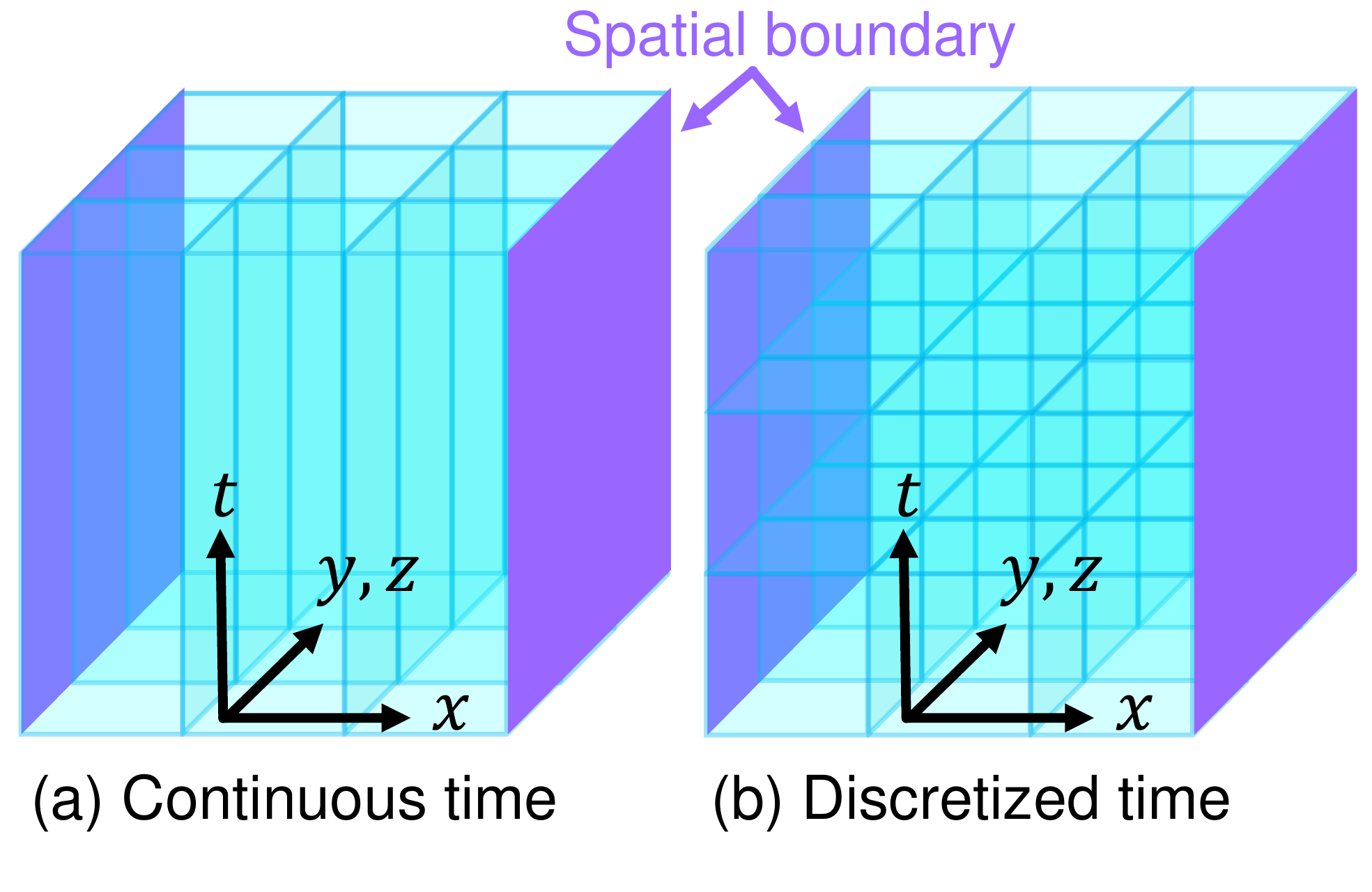}
        \end{center}
    \end{minipage}
    \caption{Two types of setup for latticized space-time, where one spatial direction has a boundary condition.
(a) The time is not latticized.
(b) The time is latticized.}
\label{fig:temporal}
\end{figure}

Before defining the Casimir energy for fermions on lattices, we clarify our setup.
In this paper, we consider a geometry where only one spatial dimension is compactified by a boundary condition.
Then one of the spatial momentum components $p_1$ is discretized.
The other momenta stay continuous.
This is the simplest geometry inducing a Casimir energy.
Besides, we have two choices to latticize the temporal direction of the geometry or not, which are painted in (a) and (b) in Fig.~\ref{fig:temporal}.
(a) When the time is continuous, the temporal component of momentum is not affected by the lattice, which corresponds to a situation realized in a condensed matter system with a small size.
(b) In contrast, latticized time appears in lattice QCD simulations.
Then the temporal momentum is discretized in a similar manner to the discretization of the spatial momenta.
These two setups lead to similar Casimir effects, but the detail is slightly different.
In this paper, we focus on the geometry (a).

To define the Casimir energy, we need the energy-momentum dispersion relation for lattice fermions.
The dispersion relation is obtained from the Dirac operator defined in a relativistic fermion action.
With nonlatticized temporal components, the (dimensionless) energy is defined as
\begin{align}
  aE(ap)=a\sqrt{D^\dagger D}, \label{eq:def_disp}
\end{align}
where $a$ is the lattice spacing, and $D$ is the Dirac operator that includes the spatial momenta and a few parameters such as the mass but does not include the temporal momentum.
This expression correctly reflects the positions of poles in the fermion propagator in momentum space.

In the space where one spatial direction is compactified, the corresponding momentum is discretized.
The discretized momentum under the periodic and antiperiodic boundary conditions is
\begin{align}
  ap_1 &\to ap_1^\mathrm{P} (n) = \frac{2 n \pi}{N}, \\
  ap_1 &\to ap_1^\mathrm{AP} (n) = \frac{(2n+1) \pi}{N},
\end{align}
respectively, and $N$ is the lattice size.
The label $n$ is an integer bounded by the (first) Brillouin zone: $0 \leq ap_k < 2\pi$ (BZ1) or $-\pi < ap_k  \leq \pi$ (BZ2).
Then the range of $n$ in both the boundary conditions is simply
\begin{align}
 0 \leq& n^{\mathrm{P,AP}} < N \ \ \ (\mathrm{BZ1}), \\
 -\frac{N}{2} <& n^{\mathrm{P,AP}} \leq \frac{N}{2} \ \ \ (\mathrm{BZ2}).
\end{align}
Note that our results of the Casimir energy on the lattice do not depend on the choice of the Brillouin zone because the Casimir energy in this paper is defined within the first Brillouin zone, and the momentum of fermions is periodic within the first Brillouin zone.

With the discretized momentum, the zero-point energy (per area) is defined as the momentum integral of the dispersion relations:
\begin{align}
  aE_0(N\to\infty) &= -N c_\mathrm{deg} \int_\mathrm{BZ} \frac{d^3ap}{(2\pi)^3} aE (ap) \nonumber\\
  \to aE_0(N) &= - c_\mathrm{deg} \int_\mathrm{BZ} \frac{d^2ap_\perp}{(2\pi)^2} \sum_n aE (ap_\perp,ap_1(n)), \label{eq:zeropoint_ene}
\end{align}
where $c_{\mathrm{deg}}$ is the degeneracy factor, including the spin degree of freedom.
For results shown in this paper, we will set $c_{\mathrm{deg}}=1$.
The subscription BZ denotes that each integration variable $p_k$ runs over the (first) Brillouin zone.
Note that the Brillouin zone for the upper (lower) equation is three- (two-) dimensional.
The negative sign in Eq.~(\ref{eq:zeropoint_ene}) is unique to fermions.
By taking account of the factor $2$ from antiparticle degrees of freedom, we drop the factor $1/2$ from the zero-point energy.

Usually, in order to obtain the Casimir energy in the continuum theory, one needs to subtract the divergent zero-point energy in infinite volume from the divergent one in finite volume.
On the other hand, in the lattice theory, both the zero-point energies are not divergent by the lattice cutoff.
In order to get a physical quantity corresponding to the Casimir energy, we give a definition on the lattice by subtracting $aE_0(N \to \infty)$ from $aE_0(N)$.
In the $3+1$-dimensional space-time with one compactified spatial dimension, the Casimir energy is~\cite{Ishikawa:2020ezm}
\begin{align}
aE_\mathrm{Cas}^\mathrm{3+1D} \equiv& aE_0(N) - aE_0(N\to\infty) \nonumber\\
=& c_\mathrm{deg} \int_\mathrm{BZ} \frac{d^2ap_\perp}{(2\pi)^2} \nonumber\\ 
& \times \left[ -\sum_n aE(ap_\perp,ap_1(n)) + N \int_\mathrm{BZ} \frac{dap_1}{2\pi} aE(ap) \right]. \label{eq:def_cas}
\end{align}
Similarly, we define the Casimir energy in lower spatial dimensions, e.g., in the $1+1$ dimensions~\cite{Ishikawa:2020ezm},
\begin{align}
aE_\mathrm{Cas}^\mathrm{1+1D} \equiv c_\mathrm{deg} \left[ -\sum_n aE(ap_1(n)) + N \int_\mathrm{BZ} \frac{dap_1}{2\pi} aE(ap) \right]. \label{eq:def_cas1+1}
\end{align}
In the following, we call the first and second terms the {\it (negative) sum part} and {\it (positive) integral part}, respectively, which is convenient to interpret the various properties of the Casimir energy.

\section{Casimir energy for naive fermion} \label{Sec_3}
In this section, we demonstrate the Casimir effect for the naive lattice fermion which is one of the simplest formulations of lattice fermions, but it contains doubler degrees of freedom.

We define the (dimensionless) Dirac operator of the naive lattice fermion in momentum space:
\begin{align}
aD_\mathrm{nf} \equiv i \sum_k \gamma_k \sin ap_k + am_f, \label{eq:nf_D}
\end{align}
where $\gamma_k$ is the gamma matrix with the index $k$, and $m_f$ is the mass of the fermion.

From the definition~(\ref{eq:def_disp}), the dispersion relation of this fermion is written as
\begin{align}
a^2 E_\mathrm{nf}^2(ap) = \sum_k \sin^2 ap_k + (am_f)^2. \label{eq:nf_E}
\end{align}
Using the square root of Eq.~(\ref{eq:nf_E}) and the definition~(\ref{eq:def_cas}), the Casimir energy in the $3+1$ dimensions is
\begin{widetext}
\begin{align}
 aE_\mathrm{Cas}^\m{3+1D,nf,P} &\equiv aE_0^\m{3+1D,nf,P}(N) - aE_0^\m{3+1D,nf,P}(N\to\infty) \\
&= c_\mathrm{deg} \int \frac{d^2ap_\perp}{(2\pi)^2} \left[ -\sum_n \sqrt{ \sin^2 \frac{2 n \pi}{N} +\sum_{k=2,3} \sin^2 ap_k + (am_f)^2} +  N \int_\mathrm{BZ} \frac{dap_1}{2\pi} \sqrt{ \sum_{k=1,2,3} \sin^2 ap_k + (am_f)^2} \right], \nonumber \\
 aE_\mathrm{Cas}^\m{3+1D,nf,AP} &\equiv aE_0^\m{3+1D,nf,AP}(N) - aE_0^\m{3+1D,nf,AP}(N\to\infty)  \\
&= c_\mathrm{deg} \int \frac{d^2ap_\perp}{(2\pi)^2} \left[ -\sum_n \sqrt{ \sin^2 \frac{(2n+1) \pi}{N} +\sum_{k=2,3} \sin^2 ap_k + (am_f)^2} + N \int_\mathrm{BZ} \frac{dap_1}{2\pi} \sqrt{ \sum_{k=1,2,3} \sin^2 ap_k + (am_f)^2}  \right], \nonumber
\end{align}
\end{widetext}
where the integration with respect to $ap_\perp$ is in the range of the first Brillouin zone.
Then we can define the Casimir energy without divergence.

\begin{figure*}[t!]
    \begin{minipage}[t]{1.0\columnwidth}
        \begin{center}
            \includegraphics[clip, width=1.0\columnwidth]{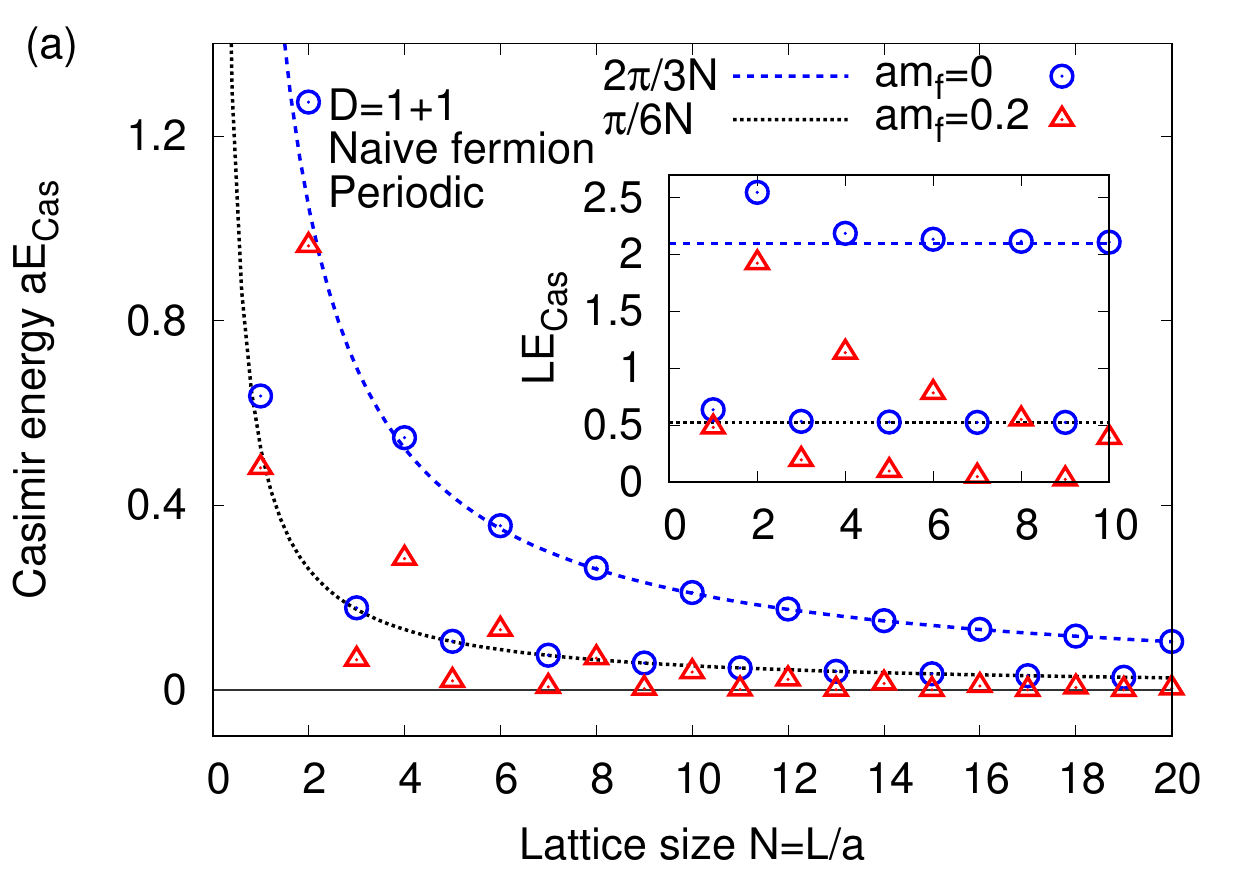}
            \includegraphics[clip, width=1.0\columnwidth]{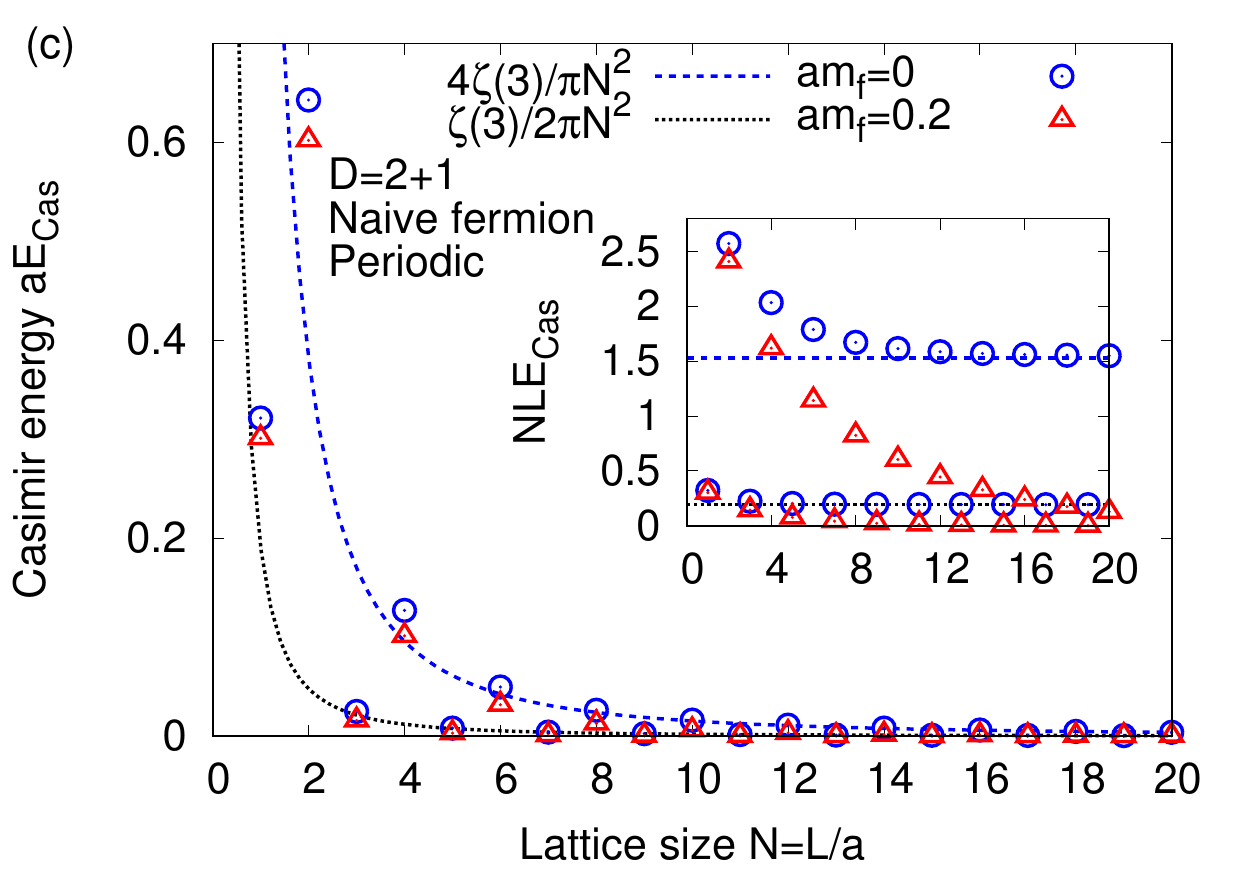}
            \includegraphics[clip, width=1.0\columnwidth]{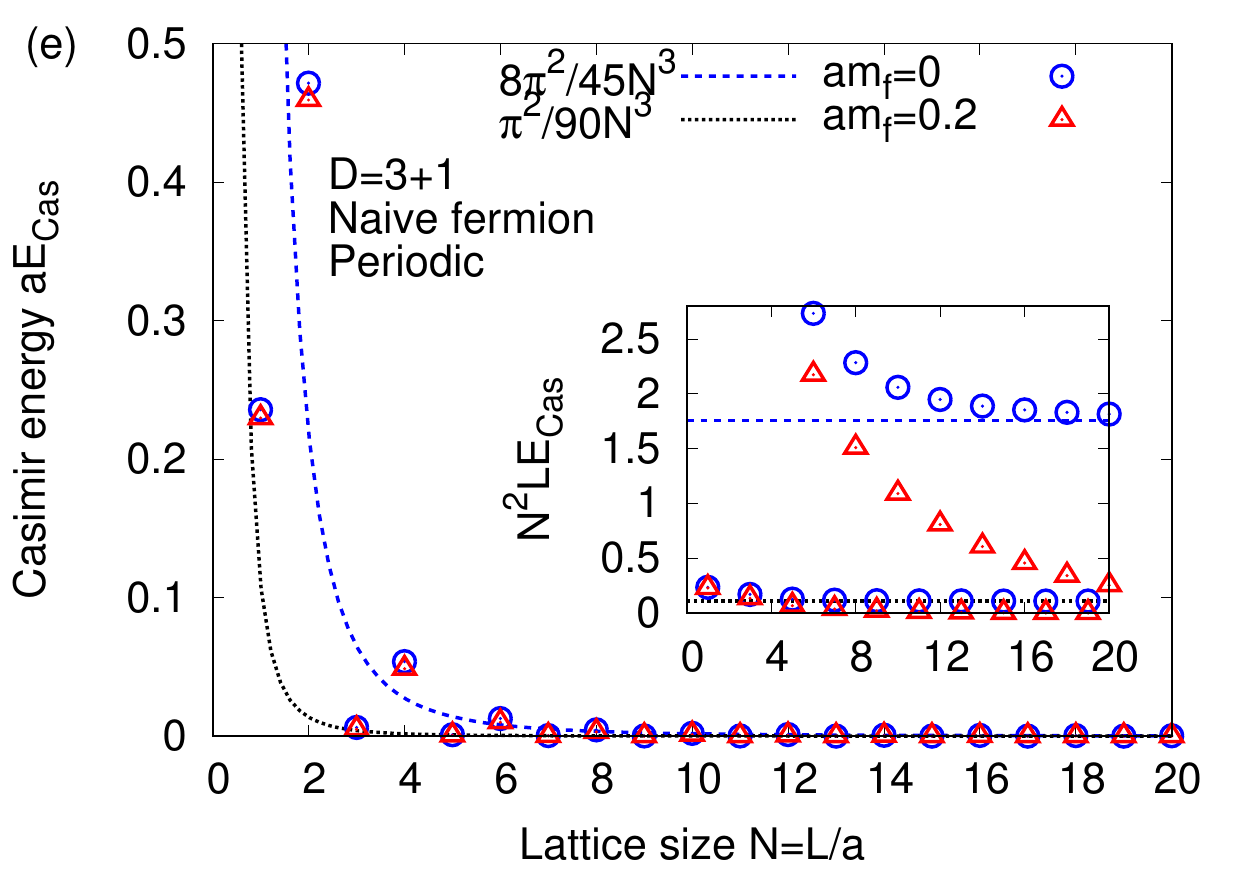}
        \end{center}
    \end{minipage}
    \begin{minipage}[t]{1.0\columnwidth}
        \begin{center}
            \includegraphics[clip, width=1.0\columnwidth]{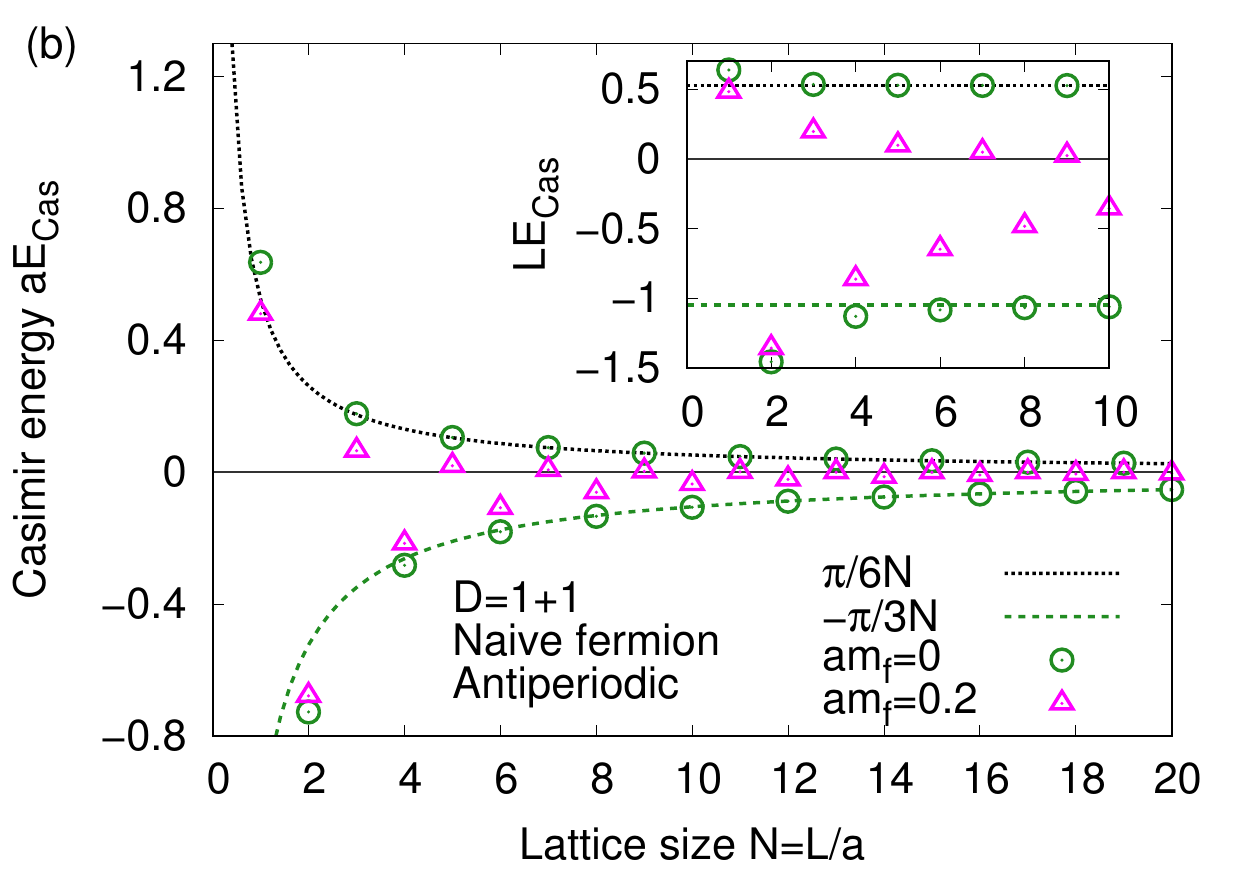}
            \includegraphics[clip, width=1.0\columnwidth]{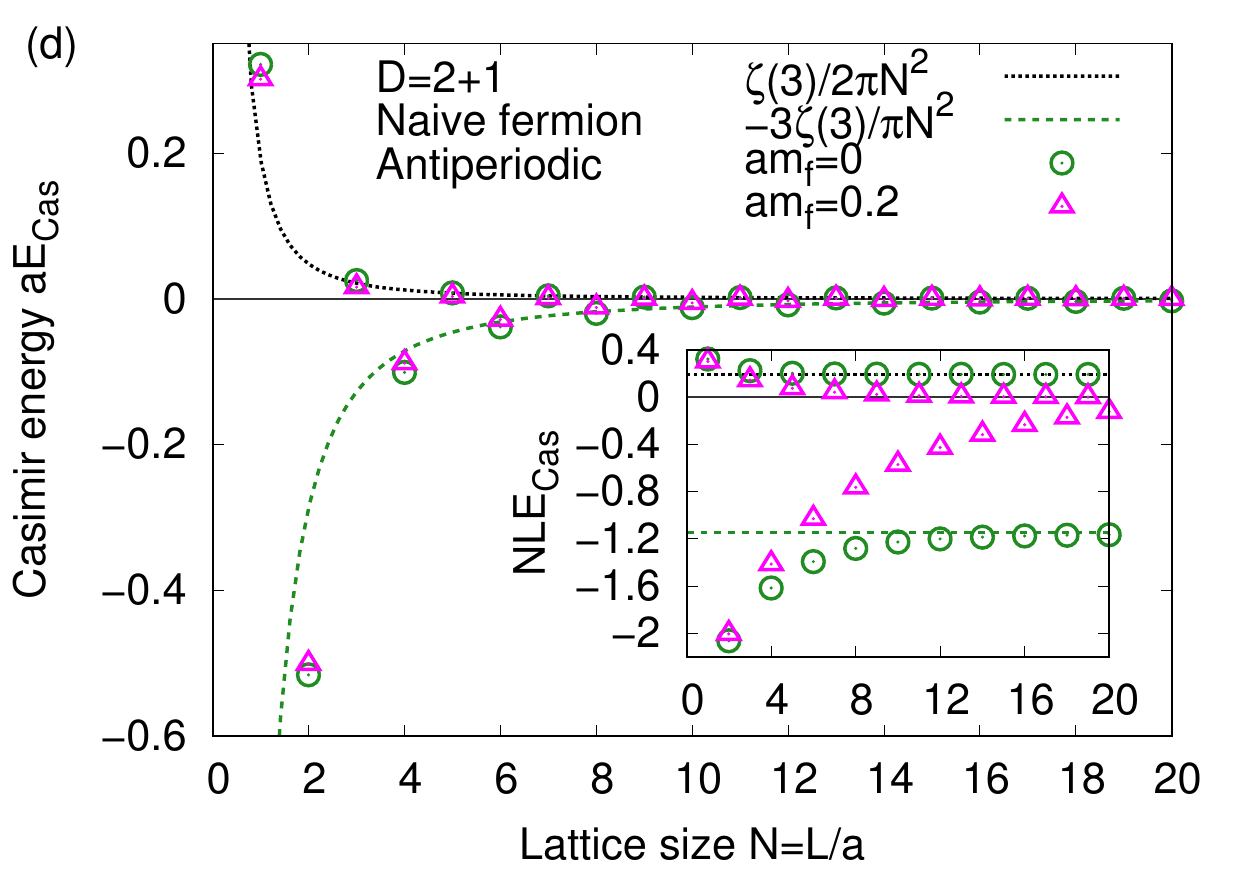}
            \includegraphics[clip, width=1.0\columnwidth]{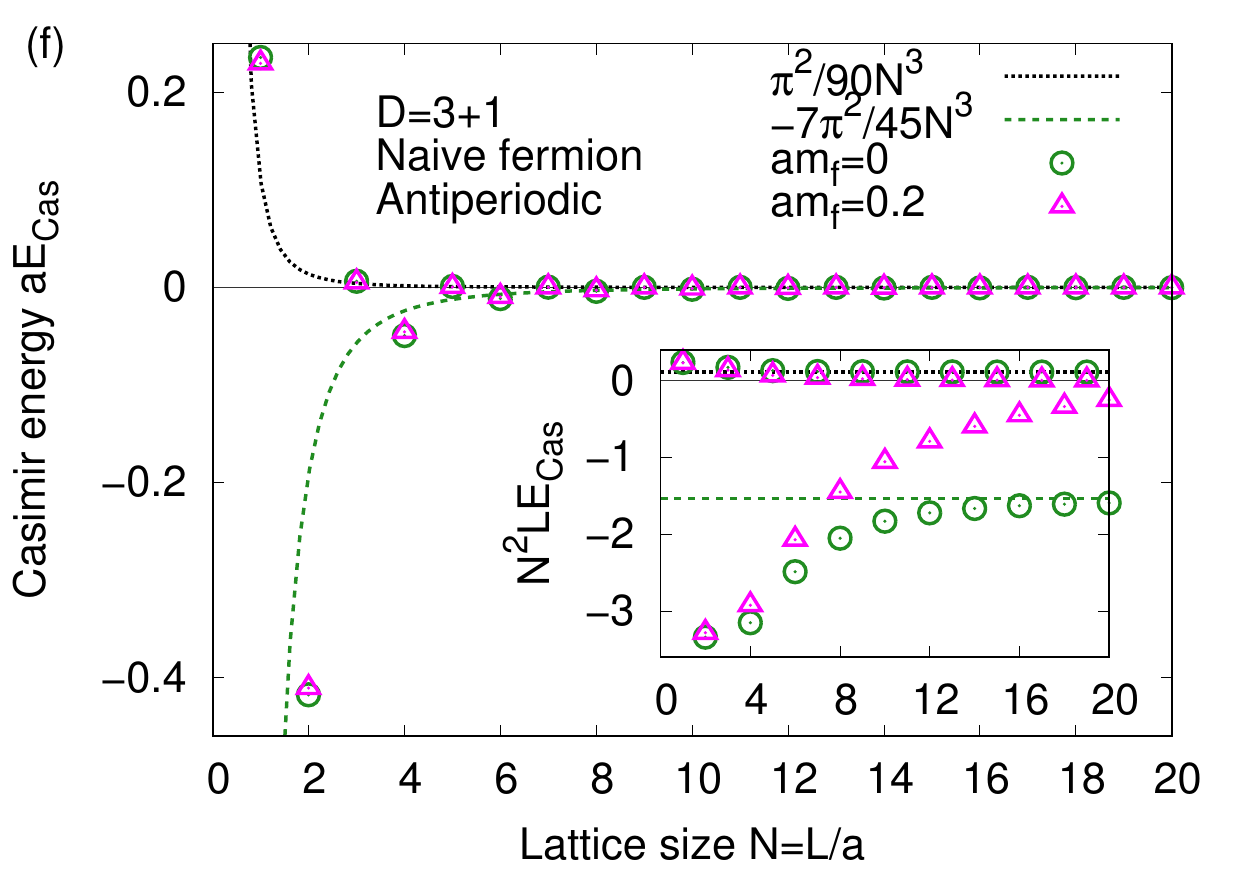}
        \end{center}
    \end{minipage}
    \caption{Casimir energy for massless or positive-mass naive fermion in the $1+1$-, $2+1$-, and $3+1$-dimensional space-time (the temporal direction is not latticized).
Small windows show the coefficients of Casimir energy.
Dashed and dotted lines are the leading terms of the expansion by $a/L$ or equivalently $1/N$, which is obtained as an asymptotic form for the massless fermion in the large lattice size $N$.
(Left) Periodic boundary.
(Right) Antiperiodic boundary.
}
\label{fig:nf}
\end{figure*}

By using mathematical techniques such as the Abel-Plana formulas (which are shown in Appendix~\ref{App:2}), one may get simpler analytic formulas for the Casimir energy.
For example, for the naive fermion with $m_f=0$ in the $1+1$-dimensional space-time, we can derive the exact formulas for dimensionless Casimir energies~\cite{Ishikawa:2020ezm} (for a derivation, see Appendix~\ref{App:3}):
\begin{align}
aE_\mathrm{Cas}^\mathrm{1+1D,nf,P} &= \frac{2N}{\pi} - \cot \frac{\pi}{2N}  & (N=\mathrm{odd}), \label{eq:nf_exact1} \\
aE_\mathrm{Cas}^\mathrm{1+1D,nf,P} &= \frac{2N}{\pi} -  2\cot \frac{\pi}{N} & (N=\mathrm{even}), \label{eq:nf_exact2} \\
aE_\mathrm{Cas}^\mathrm{1+1D,nf,AP} &= \frac{2N}{\pi} - \cot \frac{\pi}{2N} & (N=\mathrm{odd}), \label{eq:nf_exact3} \\
aE_\mathrm{Cas}^\mathrm{1+1D,nf,AP} &= \frac{2N}{\pi} - 2\csc \frac{\pi}{N} & (N=\mathrm{even}), \label{eq:nf_exact4}
\end{align}

Here, we find that the odd lattice ($N=\mathrm{odd}$) and even lattice ($N=\mathrm{even}$) exhibit different Casimir energies for both the periodic and antiperiodic boundaries.
In other words, the Casimir energy for the naive fermion is oscillatory between the even and odd lattices.
This behavior is induced by the existence of the ultraviolet-momentum zero modes (or massless doublers) in the naive fermion~\cite{Ishikawa:2020ezm}.

By expanding these formulas (\ref{eq:nf_exact1})--(\ref{eq:nf_exact4}) by $a/L$ or equivalently $1/N$, we can obtain the formulas with the small $a$ expansion~\cite{Ishikawa:2020ezm}
\begin{align}
E_\mathrm{Cas}^\mathrm{1+1D,nf,P} &= \frac{\pi}{6L} +\frac{\pi^3 a^2}{360L^3} + \mathcal{O}(a^4)   &(N=\mathrm{odd}), \label{eq:nf_aexp1} \\
E_\mathrm{Cas}^\mathrm{1+1D,nf,P} &= \frac{2\pi}{3L} +\frac{2\pi^3 a^2}{45L^3} + \mathcal{O}(a^4)  &(N=\mathrm{even}),  \\
E_\mathrm{Cas}^\mathrm{1+1D,nf,AP} &= \frac{\pi}{6L} +\frac{\pi^3 a^2}{360L^3} + \mathcal{O}(a^4)  &(N=\mathrm{odd}),  \\
E_\mathrm{Cas}^\mathrm{1+1D,nf,AP} &= -\frac{\pi}{3L} -\frac{7\pi^3 a^2}{180L^3} + \mathcal{O}(a^4)&(N=\mathrm{even}). \label{eq:nf_aexp4}
\end{align}
Thus, the terms depending on the lattice spacing $a$ are dominated by the terms with $a^2$.
From these formulas, if we take the continuum limit $a \to0 $, then the Casimir energies from the odd and even lattices are different from each other.
In other words, we cannot derive the Casimir effect for the original Dirac fermion from the continuum limit of the naive fermion formulation.
In the next sections, we will see that this situation is different from the Wilson fermion shown in Sec.~\ref{Sec_4} and the overlap fermion in Sec.~\ref{Sec_5}.

In Fig.~\ref{fig:nf}, we plot the Casimir energies for the massless or positive-mass naive lattice fermions for the periodic or antiperiodic boundary.
In these plots, it is convenient to see two types of dimensionless quantities.
The first is $a E_\mathrm{Cas}$, which is proportional to $1/N^d$ in the $d+1$ dimensions for massless particles in the continuum theory.
The second is the coefficient of Casimir energy $N^{d-1}L E_\mathrm{Cas}$, which is shown in the small windows.
This quantity is a constant in the continuum theory, so that it will be useful for comparing the difference between the lattice theory and continuum theory.
Also, in Fig.~\ref{fig:nf}, we can find that there are oscillatory behaviors of the Casimir energy in the $1+1$, $2+1$, and $3+1$ dimensions.

\begin{figure}[t!]
    \begin{minipage}[t]{1.0\columnwidth}
        \begin{center}
            \includegraphics[clip, width=1.0\columnwidth]{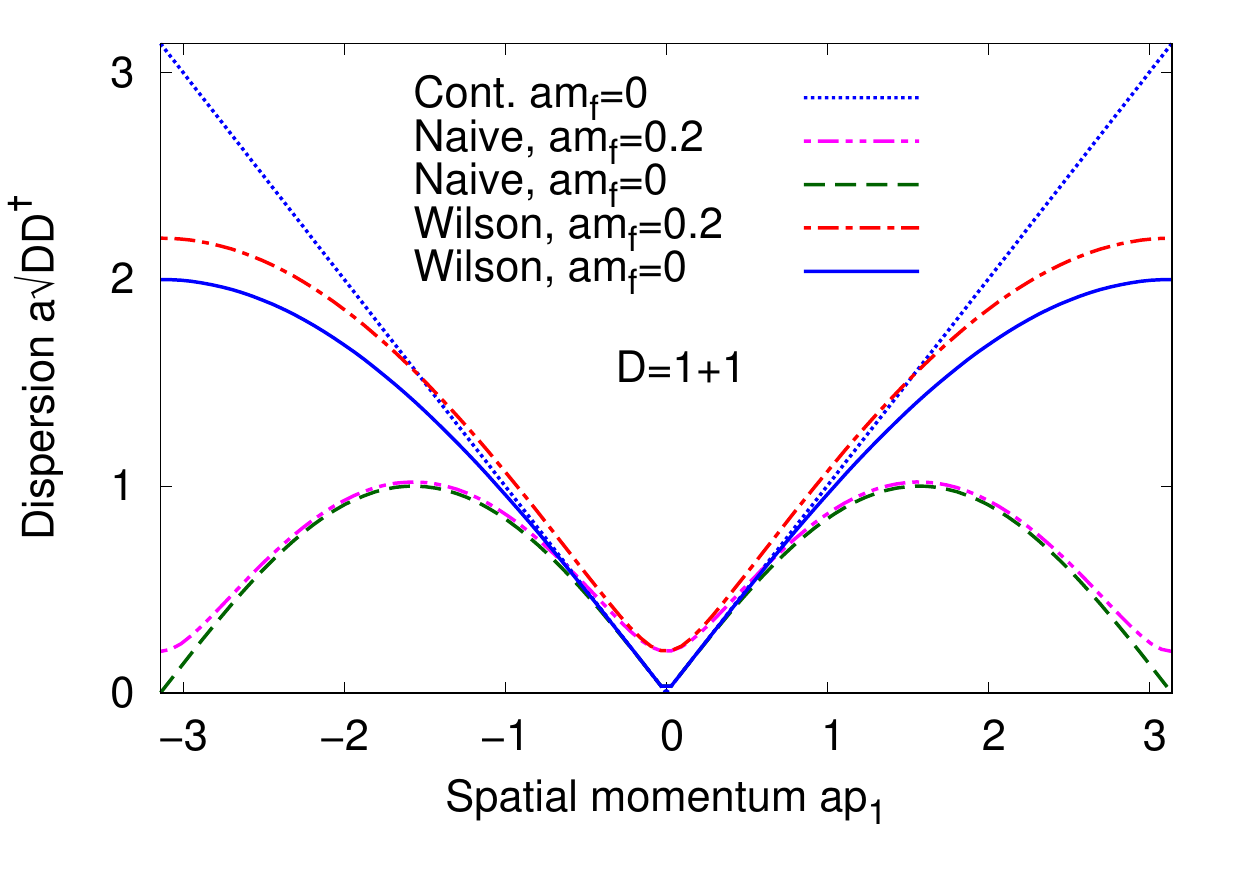}
        \end{center}
    \end{minipage}
    \caption{Dispersion relations for naive and Wilson lattice fermions in the $1+1$-dimensional space-time (the temporal direction is not latticized).}
\label{fig:2d_wil_disp}
\end{figure}

For a better understanding, it is useful to compare the Casimir energy and the corresponding dispersion relation.
In Fig.~\ref{fig:2d_wil_disp}, we show the (continuous) dispersion relations of the massless or positive-mass naive fermion in the $1+1$ dimensions, where the Brillouin zone is $-\pi< ap_1 \leq \pi$.
For the massless naive fermion, we find the dispersion relation goes to zero at $ap_1=\pi$ that is the ultraviolet-momentum zero mode as well as at $ap_1=0$ that is the infrared zero mode.
When a boundary condition is imposed, the continuous energy level is discretized into some levels living on this dispersion relation.
The Casimir energy defined as Eq.~(\ref{eq:def_cas1+1}) is determined by the difference between the sum part including contributions from these discretized levels and the integral part including the continuous level.

Furthermore, we find that the Casimir energy for the massive naive fermions is suppressed compared with that for the massless one, which is similar to the behavior of massive degrees of freedom in the continuum theory (e.g., see Refs.~\cite{Hays:1979bc,Mamaev:1980jn,Ambjorn:1981xw}), where the suppression is known to behave as an exponential function.
In fact, when we plot the logarithm of the $a E_\mathrm{Cas}$ and $N^{d-1}L E_\mathrm{Cas}$ in Fig.~\ref{fig:2d_wil_disp}, we obtain a linear behavior in the large-lattice-size region.
Therefore the exponential damping of Casimir energies for massive fermions occurs even on the lattice.
Note that the sign of the Casimir energy for a massive fermion does not change from that of the massless case.

The sign of the Casimir energy is often interesting, where positive and negative Casimir energies correspond to the repulsive and attractive Casimir forces, respectively.  
In particular, this sign is related to the number of the zero modes with $aE=0$.
\begin{itemize}
\item[(1)]{Odd $N$ with periodic boundary}---There is one infrared zero mode ($ap_1=0$).
This zero mode suppresses the contribution of the negative sum part to the Casimir energy.
As a result, the Casimir energy is dominated by the positive integral part, and its sign is positive.
\item[(2)]{Even $N$ with periodic boundary}---There are two zero modes ($ap_1=0,\pi$).
Both the zero modes suppress the negative sum part, so that the positive Casimir energy is enhanced.
\item[(3)]{Odd $N$ with antiperiodic boundary}---There is one ultraviolet-momentum zero mode ($ap_1=\pi$).
This situation is equivalent to the odd $N$ with the periodic boundary, although the momentum of the zero mode is different from each other.
Such equivalence on the odd lattice appears not only in the $1+1$ dimensions~\cite{Ishikawa:2020ezm} but also in the $2+1$ and $3+1$ dimensions.
\item[(4)]{Even $N$ with antiperiodic boundary}---In this case, there is no zero mode.
Then, the negative sum part of the Casimir energy is enhanced by the higher nonzero modes, and as a result the sign of the Casimir energy becomes negative, which is the only situation to induce the attractive Casimir force by using the free naive fermion.
\end{itemize}

\begin{figure*}[t!]
    \begin{minipage}[t]{1.0\columnwidth}
        \begin{center}
            \includegraphics[clip, width=1.0\columnwidth]{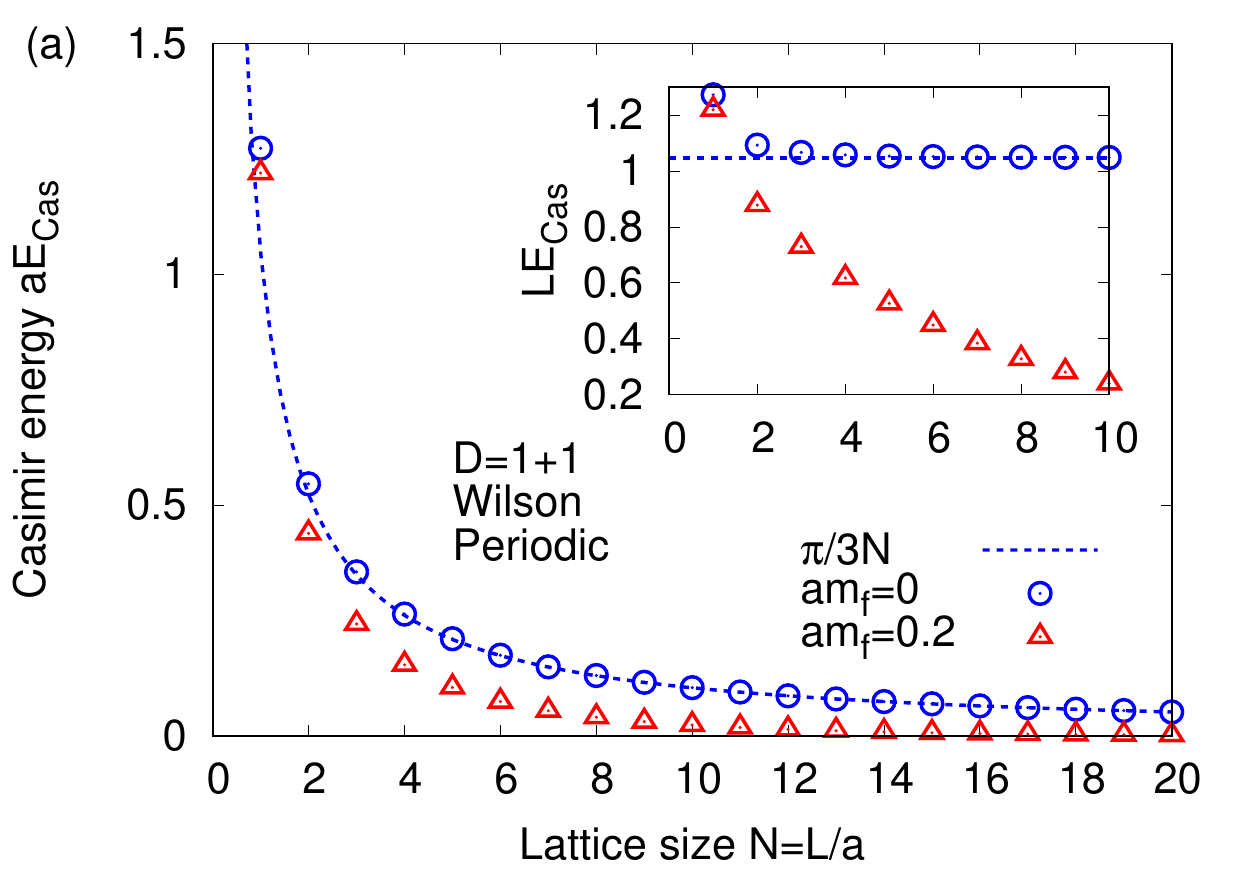}
            \includegraphics[clip, width=1.0\columnwidth]{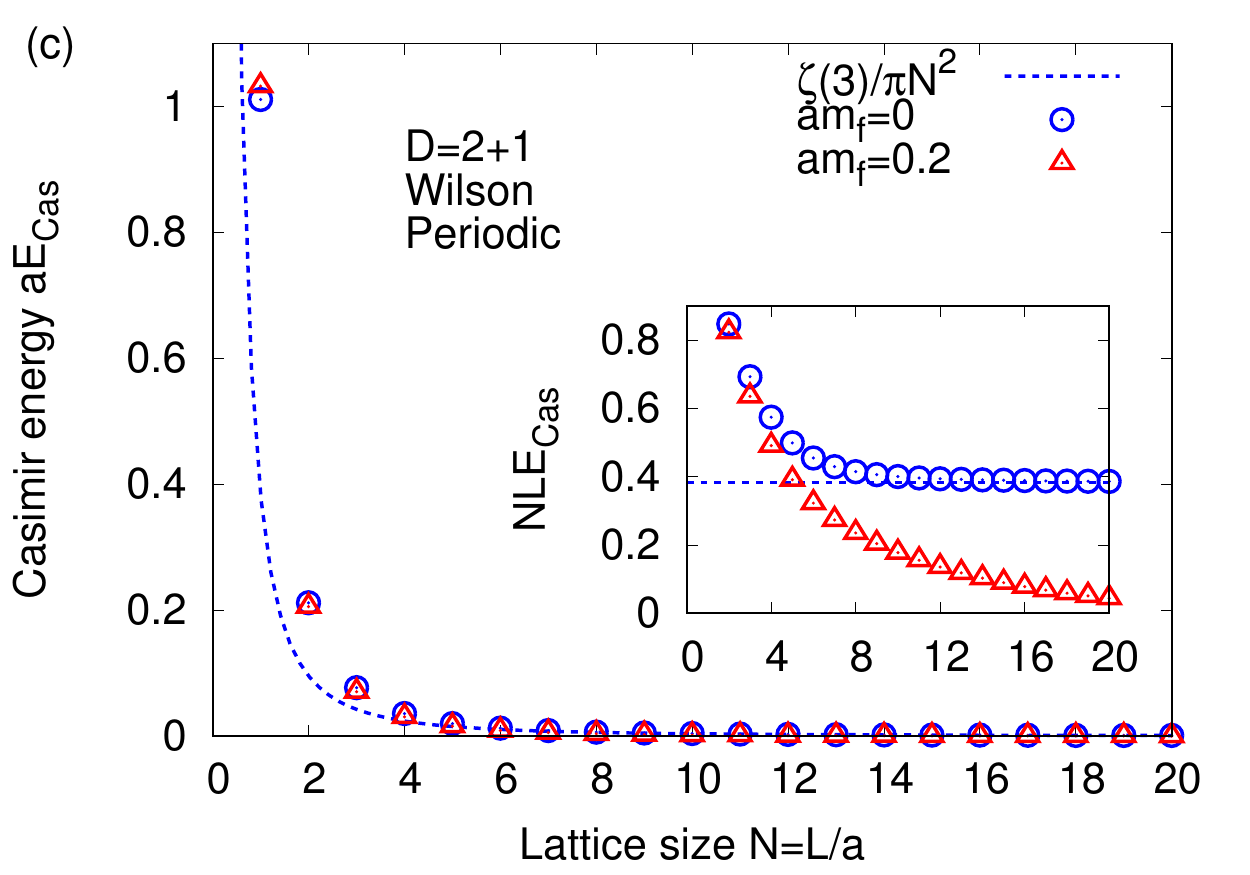}
            \includegraphics[clip, width=1.0\columnwidth]{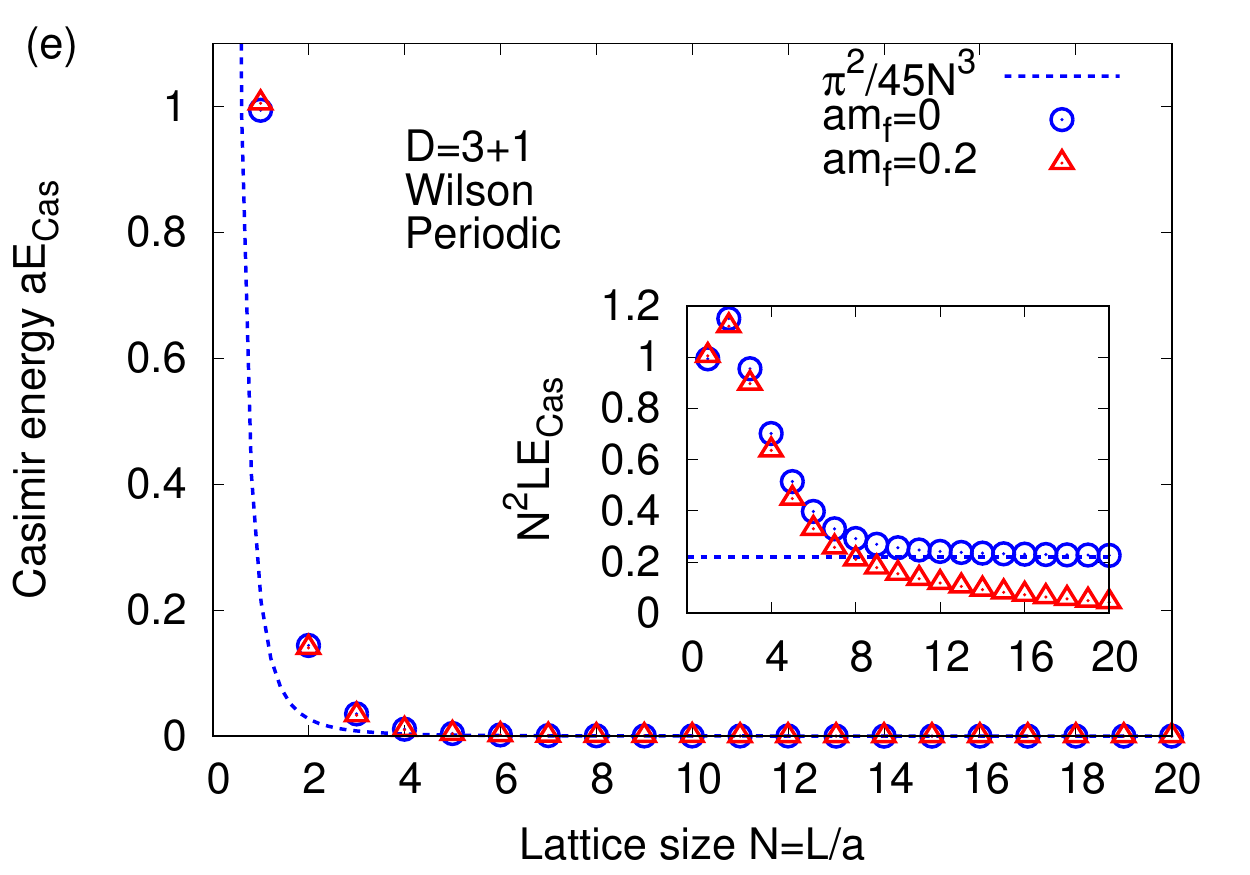}
        \end{center}
    \end{minipage}
    \begin{minipage}[t]{1.0\columnwidth}
        \begin{center}
            \includegraphics[clip, width=1.0\columnwidth]{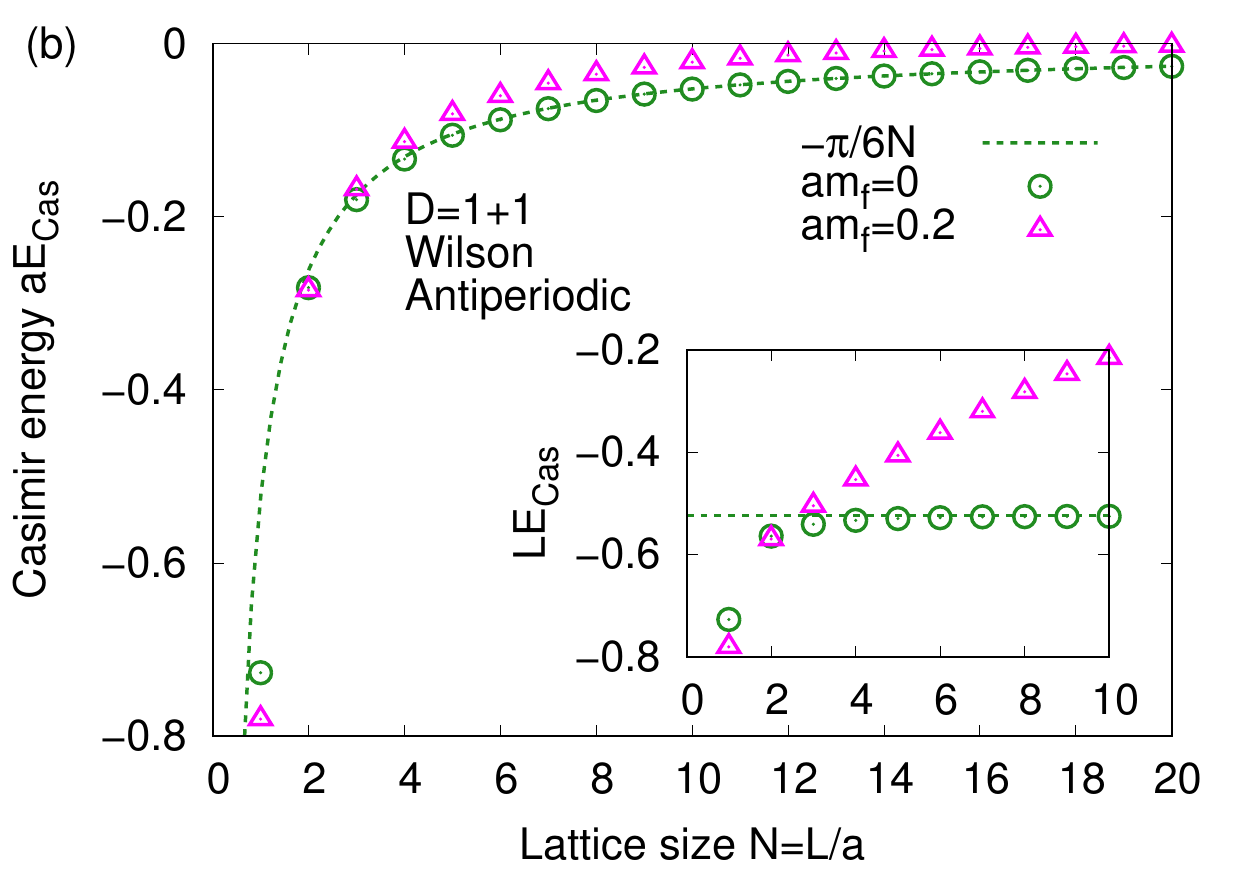}
            \includegraphics[clip, width=1.0\columnwidth]{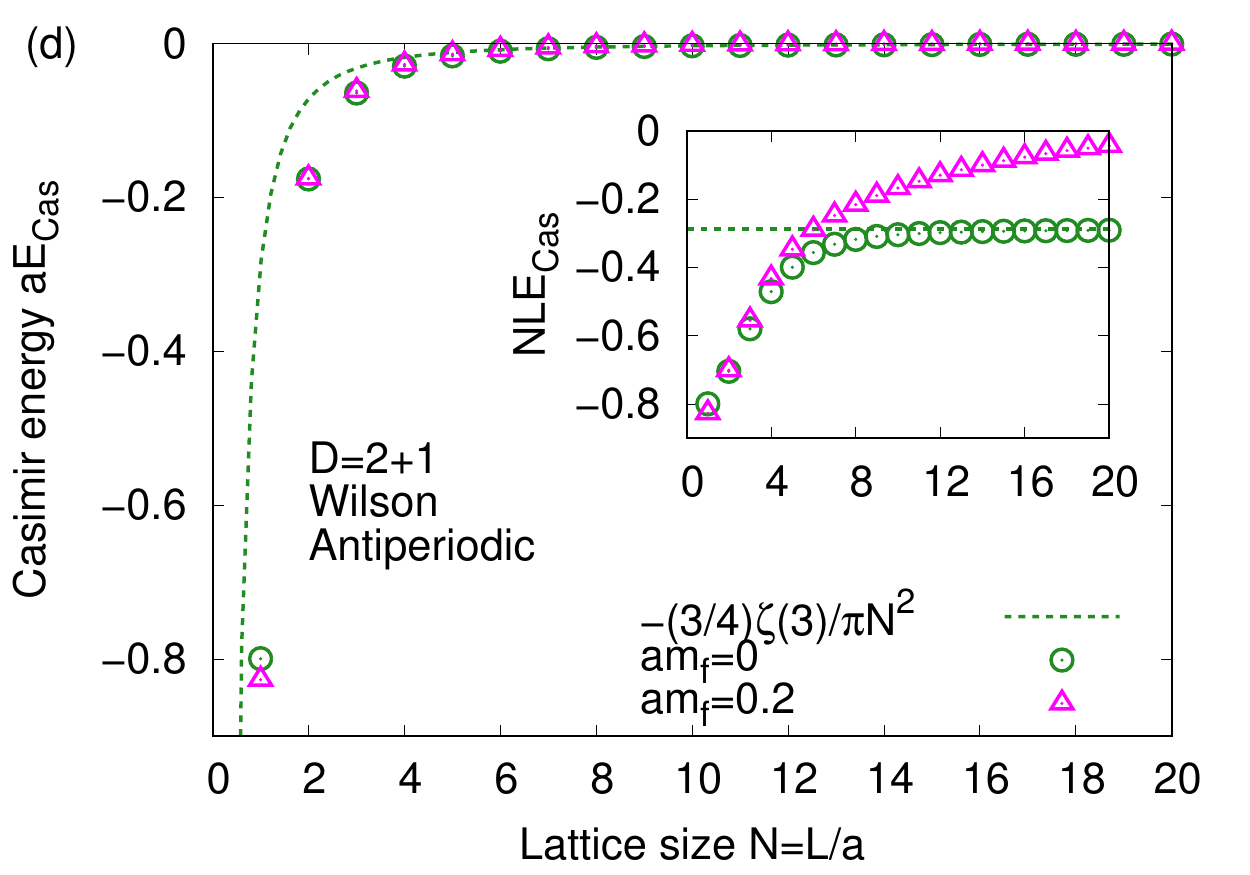}
            \includegraphics[clip, width=1.0\columnwidth]{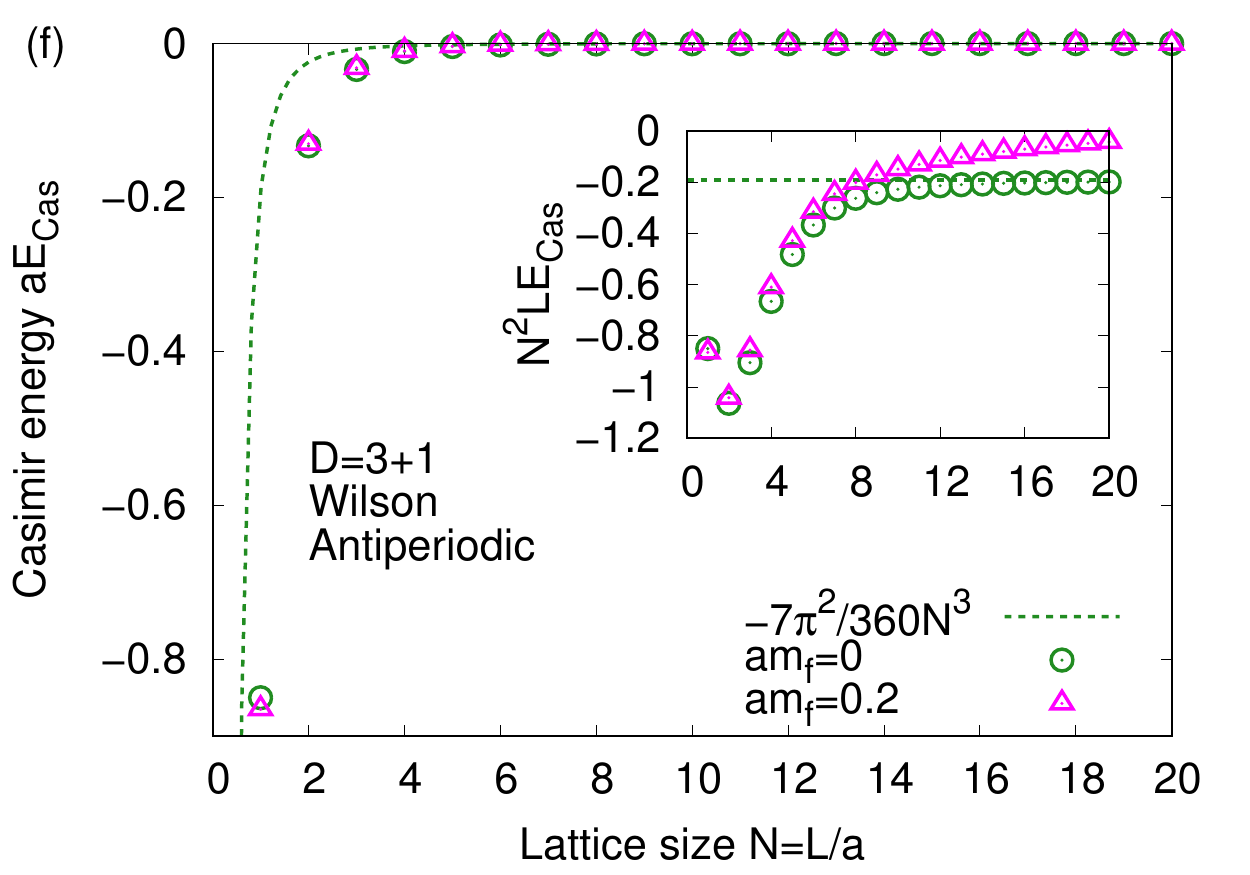}
        \end{center}
    \end{minipage}
    \caption{Casimir energy for massless or positive-mass Wilson fermion in the $1+1$-, $2+1$-, and $3+1$-dimensional space-time (the temporal direction is not latticized).
Small windows show the coefficients of Casimir energy.
Dashed lines are the leading terms of the expansion by $a/L$ or equivalently $1/N$, which is obtained as an asymptotic form for the massless fermion in the large lattice size $N$.
(Left) Periodic boundary.
(Right) Antiperiodic boundary.}
\label{fig:Wil}
\end{figure*}

\section{Casimir energy for Wilson fermion} \label{Sec_4}
In this section, we investigate the Casimir effects for Wilson fermions.
In Sec.~\ref{Sec_4-1}, we study the conventional Wilson fermion~\cite{Wilson:1975,Wilson:1977}.
The results in the $1+1$ dimensions were studied in Ref.~\cite{Ishikawa:2020ezm}, so that, in this paper, we focus on the dependence on the spatial dimension.
In Sec.~\ref{Sec_4-2}, we examine Wilson fermions with negative masses.

\subsection{Wilson fermion} \label{Sec_4-1}
The (dimensionless) Dirac operator of the Wilson fermion with the Wilson parameter $r$ and the fermion mass $m_f$ is defined as
\begin{equation}
aD_\mathrm{W} \equiv i \sum_k \gamma_k \sin ap_k + r \sum_k (1 - \cos ap_k) + am_f. \label{eq:Dw}
\end{equation}
The term proportional to $r$ is called the Wilson term, and it is interpreted as a momentum dependent mass term to eliminate the doublers.
From this Dirac operator, the dispersion relation is
\begin{equation}
a^2 E_{\mathrm{W}}^2(ap) = \sum_k \sin^2 ap_k + \left[ r \sum_k (1 - \cos ap_k) + am_f \right]^2.
\end{equation}
From the definitions of the Casimir energy, Eqs.~(\ref{eq:def_cas}) and (\ref{eq:def_cas1+1}), we can get the Casimir energy.

For example, when we set $r=1$ and $m_f=0$ in the $1+1$-dimensional space-time, we obtain the exact formulas \cite{Ishikawa:2020ezm} (for a derivation, see Appendix~\ref{App:4}):
\begin{eqnarray}
aE_\mathrm{Cas}^\mathrm{1+1D,W,P} &=& \frac{4N}{\pi} - 2\cot \frac{\pi}{2N}, \label{eq:Wil_exactP} \\
aE_\mathrm{Cas}^\mathrm{1+1D,W,AP} &=& \frac{4N}{\pi} - 2\csc \frac{\pi}{2N}. \label{eq:Wil_exactAP}
\end{eqnarray}
Expanding these formulas by $a/L$, we obtain \cite{Ishikawa:2020ezm}
\begin{eqnarray}
E_\mathrm{Cas}^\mathrm{1+1D,W,P} &=& \frac{\pi}{3L} +\frac{\pi^3 a^2}{180L^3} + \mathcal{O}(a^4)  \\
E_\mathrm{Cas}^\mathrm{1+1D,W,AP} &=& -\frac{\pi}{6L} -\frac{7\pi^3 a^2}{1440L^3} + \mathcal{O}(a^4).
\end{eqnarray}
Here, we find that the terms with $\frac{\pi}{3L}$ for the periodic boundary and $-\frac{\pi}{6L}$ for the antiperiodic boundary agree with the Casimir energy for the massless fermion in continuum theory, respectively (for a derivation, see Appendix~\ref{App:1}):
\begin{eqnarray}
E_\mathrm{Cas}^\mathrm{1+1D,cont,P} &=& \frac{\pi}{3L}, \label{eq:2dcont_PBC} \\
E_\mathrm{Cas}^\mathrm{1+1D,cont,AP} &=& -\frac{\pi}{6L}. \label{eq:2dcont_APB}
\end{eqnarray}
Thus, by taking the continuum limit $a \to 0$ of the Casimir energy for the Wilson fermion, we can obtain that for the continuum theory.
This is because the Wilson fermion has no doublers in contrast to the naive fermion, and also the Casimir energy is well dominated by the infrared zero modes.
On the other hand, the terms including $a^2$ and the higher orders are lattice effects (or ``lattice artifacts" in the context of lattice simulations) which is related to properties in the ultraviolet region of the dispersion relation.

In Fig.~\ref{fig:Wil}, we plot the Casimir energy for the massless or positive-mass Wilson fermion at $r=1$.
We find that, even for the $2+1$- and $3+1$- dimensional space-time, the Casimir energies for the massless Wilson fermions are agree well with the continuum limit:
\begin{eqnarray}
E_\mathrm{Cas}^\mathrm{2+1D,cont,P} &=& \frac{\zeta(3)}{\pi L^2}, \label{eq:3dcont_PBC} \\
E_\mathrm{Cas}^\mathrm{2+1D,cont,AP} &=& -\frac{3\zeta(3)}{4\pi L^2}, \label{eq:3dcont_APB} \\
E_\mathrm{Cas}^\mathrm{3+1D,cont,P} &=& \frac{\pi^2}{45 L^3}, \label{eq:4dcont_PBC} \\
E_\mathrm{Cas}^\mathrm{3+1D,cont,AP} &=& -\frac{7 \pi^2}{360L^3}. \label{eq:4dcont_APB}
\end{eqnarray}
Also, we find that the Casimir energy for the positive-mass Wilson fermions is suppressed compared with that for the massless one, which is similar to the behavior of massive naive fermions in Sec.~\ref{Sec_3}.

Finally, we discuss the relation between the sign of the Casimir energy and the number of the zero modes:
\begin{itemize}
\item[(1)]{Periodic boundary}---For the massless case, there is one infrared zero mode ($ap_1=0$).
This zero mode suppresses the negative sum part to the Casimir energy, so that the Casimir energy becomes positive.
When the fermion has a positive mass and the zero mode disappears, both the negative sum part and positive integral part are enhanced.
As a result, the negative Casimir energy is suppressed.
\item[(2)]{Antiperiodic boundary}---Even for the massless case, there is no zero mode.
Then, the negative sum part of the Casimir energy is enhanced by the higher nonzero modes, so that the Casimir energy is negative.
\end{itemize}

\subsection{Wilson fermions with a negative mass} \label{Sec_4-2}
Next, we study the Casimir energy for the Wilson fermions with a negative mass, $am_f<0$.
A negative mass in the Wilson fermion corresponds to a negative shift of the Wilson term, and the resultant dispersion relation is significantly modified.
This situation is distinct from the effect of a negative mass in the continuous Dirac fermion.
In the Dirac fermion, the sign of the mass term in the Lagrangian does not affect the dispersion relations such as $E=\pm \sqrt{p^2+m^2}$.

In realistic topological insulators, the parameter $am_f$ (and the Wilson parameter $r$) is related to the strength of the spin-orbit interaction and the original band structure without the spin-orbit interaction, which are intrinsic to a material.
In experiments, one can tune this parameter by changing the chemical composition of the material (for example, for BiTl(S${}_{1-\delta}$Se${}_{\delta})_2$, see Ref.~\cite{Xu:2011dc}).

\begin{figure*}[t!]
    \begin{minipage}[t]{1.0\columnwidth}
        \begin{center}
            \includegraphics[clip, width=1.0\columnwidth]{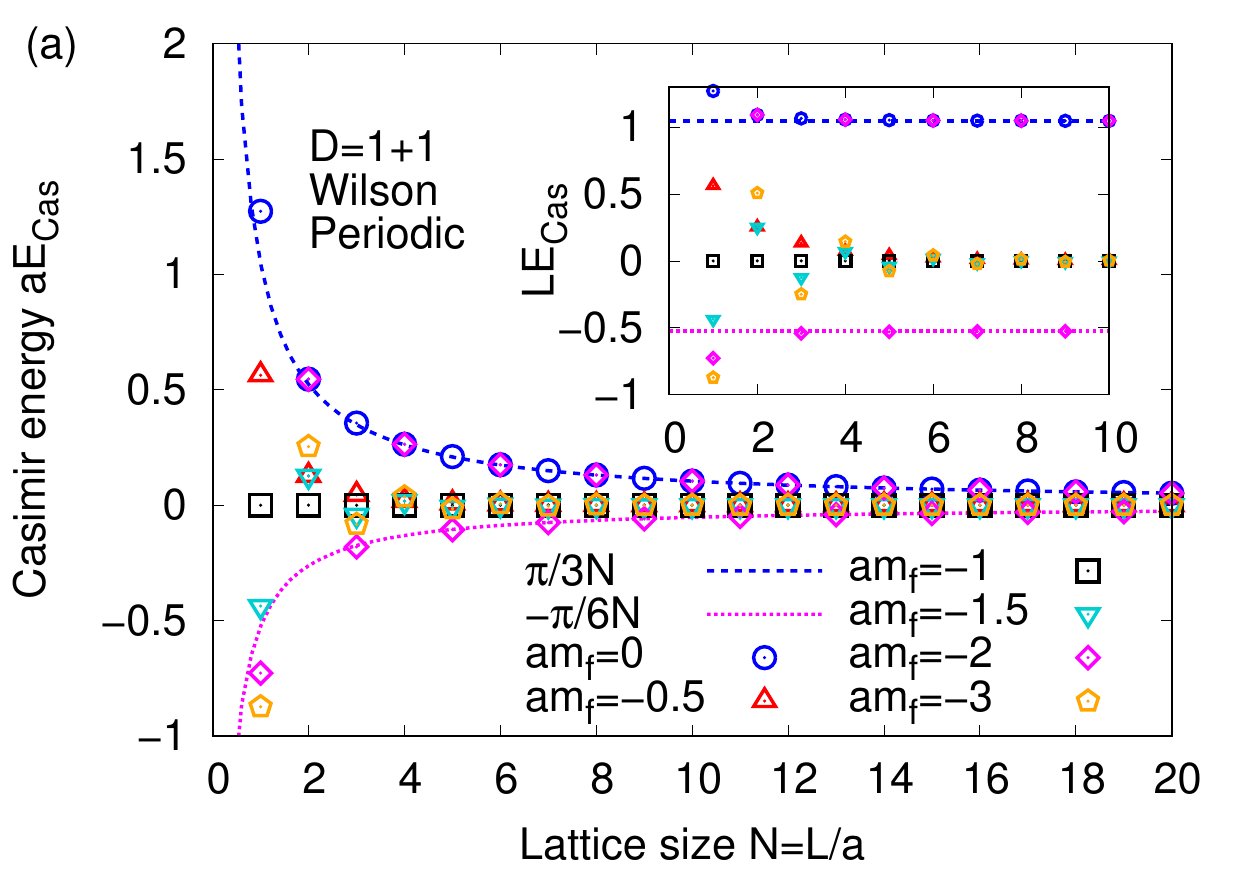}
            \includegraphics[clip, width=1.0\columnwidth]{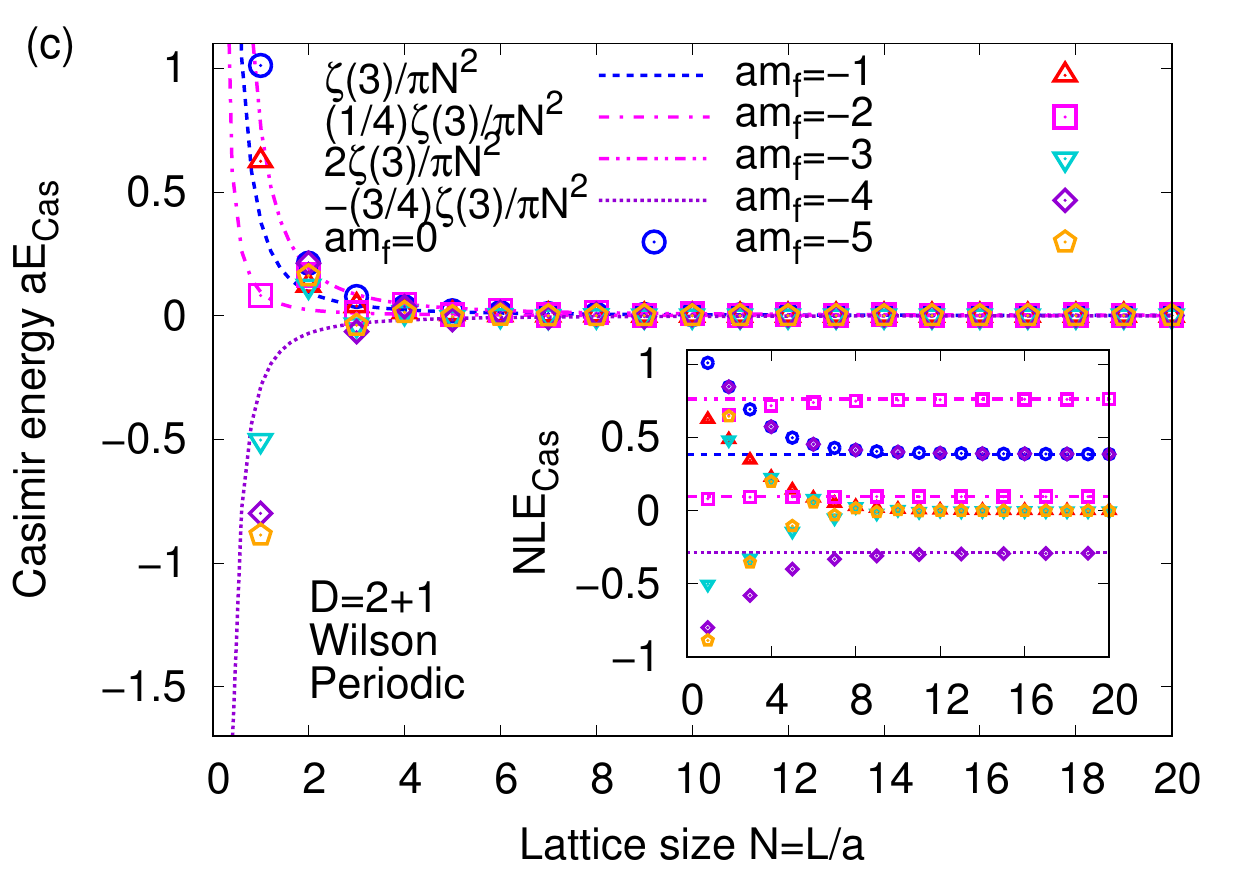}
        \end{center}
    \end{minipage}
    \begin{minipage}[t]{1.0\columnwidth}
        \begin{center}
            \includegraphics[clip, width=1.0\columnwidth]{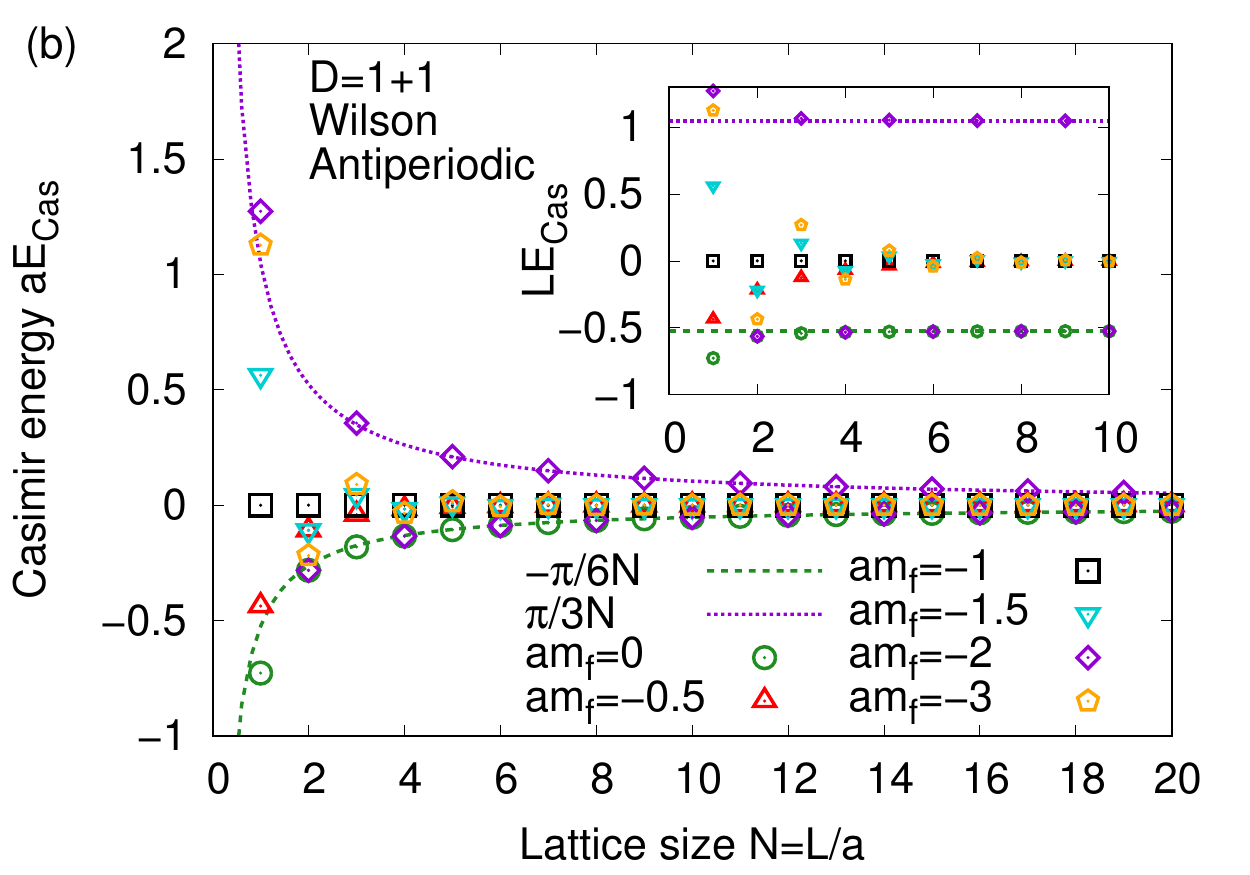}
            \includegraphics[clip, width=1.0\columnwidth]{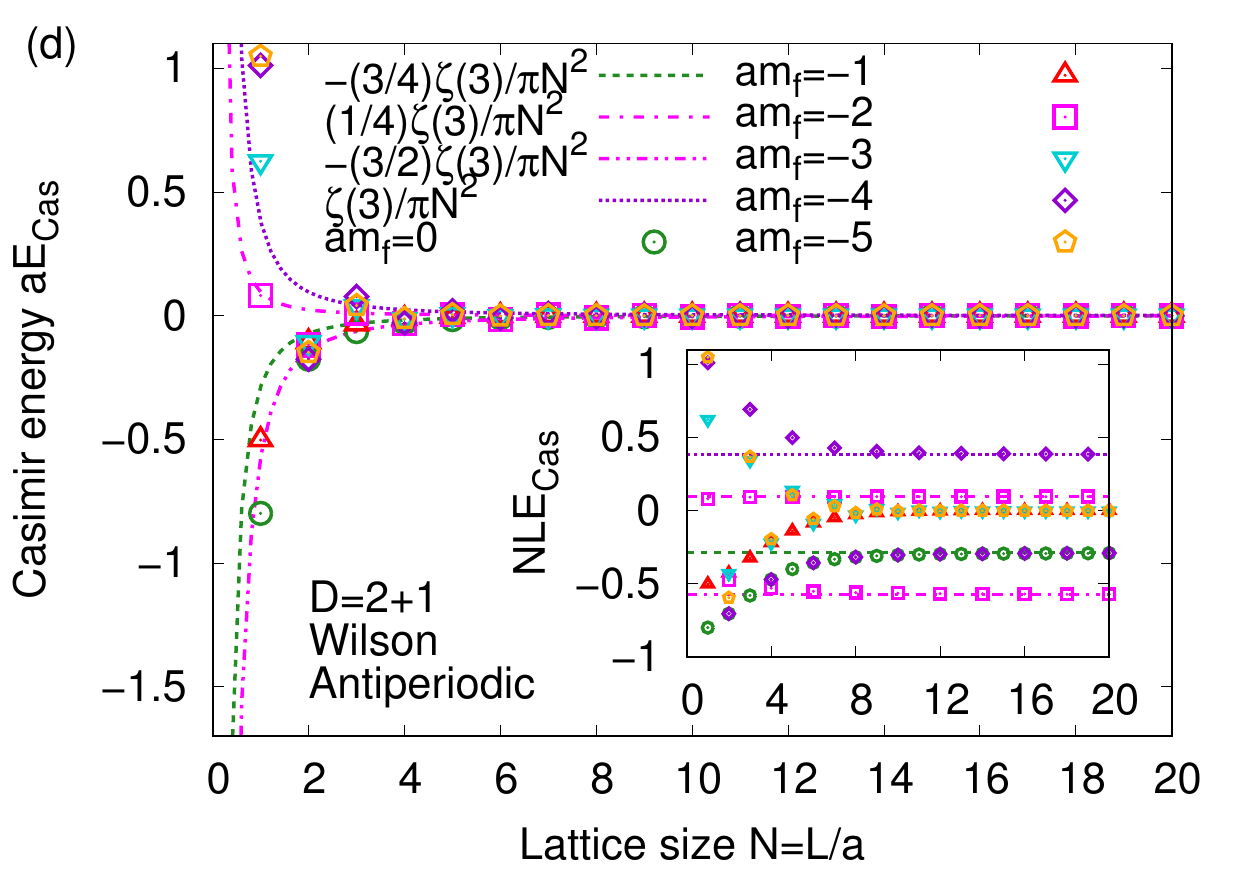}
        \end{center}
    \end{minipage}
    \caption{Casimir energy for Wilson fermions with a negative mass in the $1+1$- and $2+1$-dimensional space-time (the temporal direction is not latticized).
Small windows show the coefficients of Casimir energy.
Dashed, dotted, and dash-dotted lines are the leading terms of the expansion by $a/L$ or equivalently $1/N$, which is obtained as an asymptotic form in the large lattice size $N$.
(Left) Periodic boundary.
(Right) Antiperiodic boundary.}
\label{fig:Wil_nega}
\end{figure*}

\begin{figure}[t!]
    \begin{minipage}[t]{1.0\columnwidth}
        \begin{center}
            \includegraphics[clip, width=1.0\columnwidth]{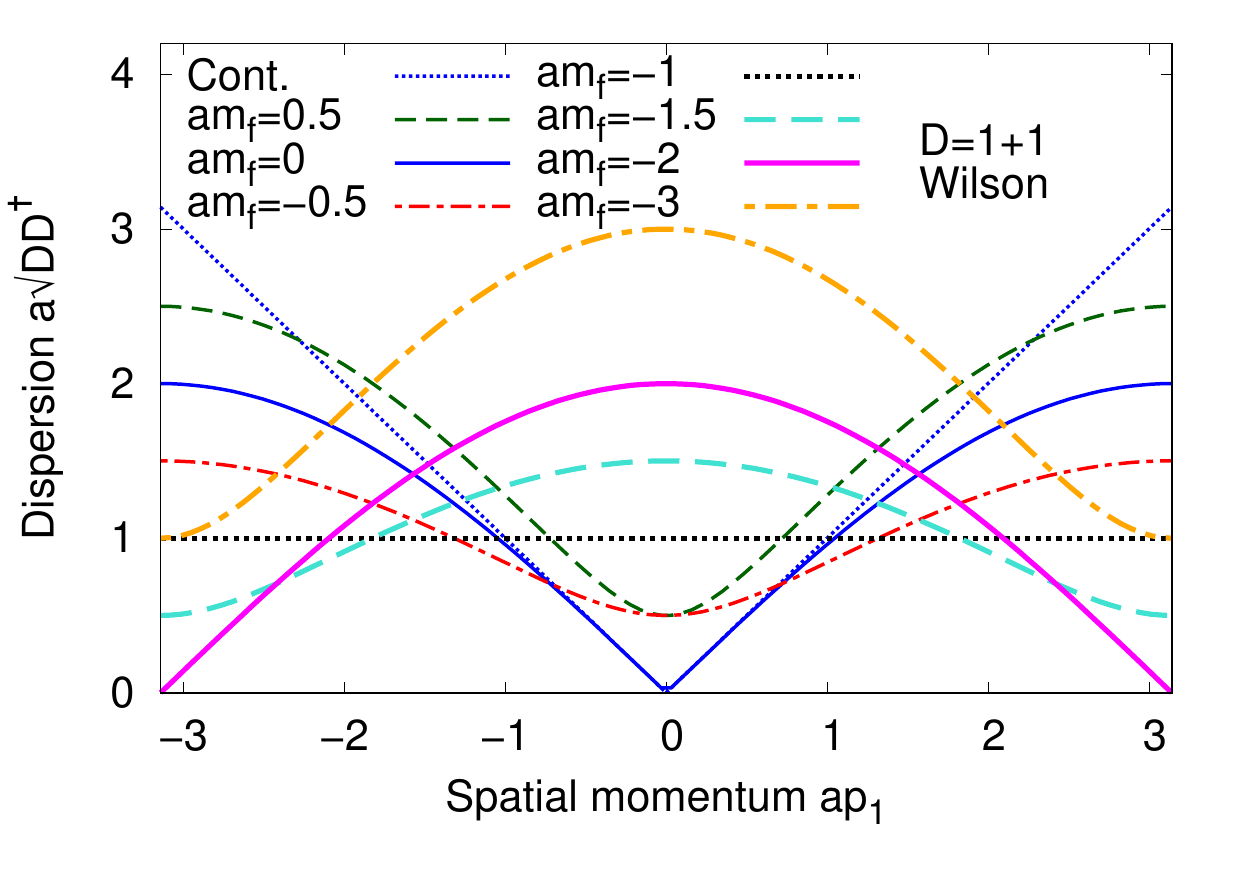}
        \end{center}
    \end{minipage}
    \caption{Negative-mass dependence of dispersion relations for Wilson fermions in the $1+1$-dimensional space-time (the temporal direction is not latticized).}
\label{fig:2d_Wil_neg_disp}
\end{figure}

In the upper panels of Fig.~\ref{fig:Wil_nega}, we show the numerical results in the $1+1$-dimensional space-time, where the spatial one dimension has the periodic or antiperiodic boundary condition.
The properties of the Casimir energy can be well understood by their dispersion relations, as shown in Fig.~\ref{fig:2d_Wil_neg_disp}.
From Fig.~\ref{fig:2d_Wil_neg_disp}, we can see that the energy of the low-momentum mode increases by a negative mass, while that of the high-momentum mode decreases in the region of $0 < am_f < -2$.
Here, our findings are as follows:
\begin{itemize}
\item[(1)]{\it $-1 < am_f \leq 0$}---At $am_f=0$, we observe the Casimir energy of massless Wilson fermions.
As $am_f$ decreases, both the Casimir energy $aE_\mathrm{Cas}$ and its coefficient $LE_\mathrm{Cas}$ is suppressed in larger lattice size, which is similar to that for Wilson fermions with a positive mass.
This similarity is understood from the dispersion relations shown in Fig.~\ref{fig:2d_Wil_neg_disp}.
For example, the dispersion for the negative mass with $am_f=-0.5$ (the red dash-dotted curve) is similar to that for the positive mass $am_f=0.5$ (the green dashed curve).
\item[(2)]{\it $am_f=-1$}---At this parameter, we find that the Casimir energy is equal to zero in both the periodic and antiperiodic boundaries.
This behavior can be understood as follows.
The Wilson Dirac operator at $am_f=-1$ is written as
\begin{align}
aD_{\mathrm{1+1D}} = i \gamma_1 \sin{ap_1} + r \left( - \cos ap_1 \right). \label{eq:Dw_am=-1}
\end{align}
Then, we find that the dispersion relation is a constant, as shown in the black dotted line of Fig.~\ref{fig:2d_Wil_neg_disp}:
\begin{align}
a^2 E_{\mathrm{1+1D}}^2(ap) &= 1. \label{eq:E_am=-1}
\end{align}
Using the definition (\ref{eq:def_cas1+1}), we can show that the corresponding Casimir energy is zero:
\begin{align}
aE_\mathrm{Cas}^\mathrm{1+1D} &= 0. \label{eq:ECas_am=-1}
\end{align}
Thus, $am_f=-1$ is a special mass parameter for the Casimir effect in the $1+1$ dimensions.

The physical interpretation of this behavior is as follows: 
For $am_f=-1$, the dispersion relation of the Wilson fermion becomes a ``flat band" where the energy is independent of the momentum (for an equivalent discussion, see Ref.~\cite{Junemann:2016fxu}).
The Casimir effect originates from the difference between the zero point energies of the finite volume and infinite volume (or finite lattice size $N \neq \infty$ and infinite lattice size $N = \infty$),
but the energy difference for the flat band is zero, which means that the Casimir effect does not occur.
As an alternative interpretation, this flat band means that its eigenstates in real space are wave functions ``localized" on 
nearest-neighbor two sites, where a particle can hop only between the two sites and does not hop onto other sites.
For such localized wave functions, when we change the numbers of lattices, the energy (per one site) of the systems does not change.

\item[(3)]{\it $-2<am_f<-1$}---In this region, we find an oscillation of Casimir energy between the even and odd lattices, and $LE_\mathrm{Cas}$ is suppressed in larger lattice size.
As shown in the cyan dashed curve of Fig.~\ref{fig:2d_Wil_neg_disp}, this oscillation is dominated by lower nonzero modes around $ap_1=\pi$, and the contribution from higher modes around $ap_1=0$ is relatively suppressed.
\item[(4)]{\it $am_f=-2$}---This is also a special parameter, but it is distinct from $am_f=-1$.
Here, we find an oscillation of the Casimir energy, and $LE_\mathrm{Cas}$ approaches to a constant in larger lattice size, which implies that the Casimir energy is dominated by massless degrees of freedom.

The Wilson Dirac operator at $am_f=-2$ is written as
\begin{align}
aD_{\mathrm{1+1D}} = i \gamma_1 \sin{ap_1} + r \left(-1 - \cos ap_1 \right). \label{eq:Dw_am=-2}
\end{align}
Then, we find that the dispersion has a ultraviolet-momentum zero mode (in other words, massless doubler) at $ap_1=\pi$, as shown in the magenta solid curve of Fig.~\ref{fig:2d_Wil_neg_disp}.
Note that Fig.~\ref{fig:2d_Wil_neg_disp} is just a plot of continuous spectrum.
Although energy levels discretized by the existence of boundaries is not necessarily to pick up this zero mode, which depends on the form of the boundary condition, at least the Casimir energy is dominated by ``light" modes around this zero mode.
The exact formulas of the Casimir energies at $am_f=-2$ and $r=1$ are (for a derivation, see Appendix~\ref{App:5})
\begin{eqnarray}
aE_\mathrm{Cas}^\mathrm{1+1D,P} &=& \frac{4N}{\pi} - 2\csc \frac{\pi}{2N}  \ \ (N=\mathrm{odd}), \label{eq:Wil_m=-2_exact1} \\
aE_\mathrm{Cas}^\mathrm{1+1D,P} &=& \frac{4N}{\pi} - 2\cot \frac{\pi}{2N} \ \ (N=\mathrm{even}), \label{eq:Wil_m=-2_exact2}\\
aE_\mathrm{Cas}^\mathrm{1+1D,AP} &=& \frac{4N}{\pi} - 2\cot \frac{\pi}{2N} \ \ (N=\mathrm{odd}), \label{eq:Wil_m=-2_exact3}\\
aE_\mathrm{Cas}^\mathrm{1+1D,AP} &=& \frac{4N}{\pi} - 2\csc \frac{\pi}{2N} \ \ (N=\mathrm{even}). \label{eq:Wil_m=-2_exact4}
\end{eqnarray}
Here, we find that Eqs.~(\ref{eq:Wil_m=-2_exact2}) and (\ref{eq:Wil_m=-2_exact3}) agree with Eq.~(\ref{eq:Wil_exactP}) for the periodic boundary at $am_f=0$.
On the other hand, Eqs.~(\ref{eq:Wil_m=-2_exact1}) and (\ref{eq:Wil_m=-2_exact4}) agree with Eq.~(\ref{eq:Wil_exactAP}) for the antiperiodic boundary at $am_f=0$.

Thus we find that, on the even lattice, {\it the Casimir energies for the $am_f=0$ and $am_f=-2$ are equivalent to each other.}
This is because, for the periodic boundary, the contribution from the infrared zero mode at $am_f=0$ is equivalent to that from the ultraviolet-momentum zero mode at $am_f=-2$, which holds also for contribution from nonzero modes.
The existence of zero modes leads to the positive Casimir energy.
For the antiperiodic boundary, the dispersion does not contain zero modes, so that it induces negative Casimir energy.

On the odd lattice, {\it the Casimir energy with the periodic (antiperiodic) boundary at $am_f=-2$ is equivalent to that with the antiperiodic (periodic) boundary at $am_f=0$.}
This is because the infrared zero mode at $am_f=0$ for the periodic boundary plays an equivalent role of the ultraviolet-momentum zero mode at $am_f=-2$ for the antiperiodic boundary, where their zero modes induce the positive Casimir energy.
Also, the dispersions at $am_f=0$ for the antiperiodic boundary and at $am_f=-2$ for the periodic boundary do not include zero modes, so that it contributes to negative Casimir energy.
\item[(5)]{\it $am_f<-2$}---In this region, we also find an oscillation between the odd and even lattices, where $LE_\mathrm{Cas}$ is suppressed in larger lattice size.
Such behavior is similar to the region of $-2<am_f<-1$.
\end{itemize}

Here, we comment on the relationship between our results and 1D topological insulators which can be described by the Su-Schrieffer-Heeger (SSH) model~\cite{Su:1979ua,Su:1979wc} as a typical model.
While the Wilson fermion with either a positive mass $am_f>0$ or a large negative mass $am_f<-2$ is a normal insulator without nontrivial topology, systems described by the Wilson fermion with $-2<am_f<0$ corresponds to a topological insulator, where the nontrivial topology is characterized by the Winding number $\nu=1$ corresponding to the mapping $S^1 \to S^1$ from momentum space to spin space.
The existence of an open boundary condition (or a finite length) induces a gapless chiral edge mode at the zero-dimensional edge of the insulator.
The Casimir effect for bulk modes inside the topological insulator will affect the thermodynamic properties of the bulk modes.
Note that, as shown in Eq.~(\ref{eq:E_am=-1}) and the black dotted line in Fig.~\ref{fig:2d_Wil_neg_disp}, $am_f=-1$ is a special parameter, which corresponds to just a flat band without any momentum dependence.

\begin{figure}[t!]
    \begin{minipage}[t]{1.0\columnwidth}
        \begin{center}
            \includegraphics[clip, width=0.45\columnwidth]{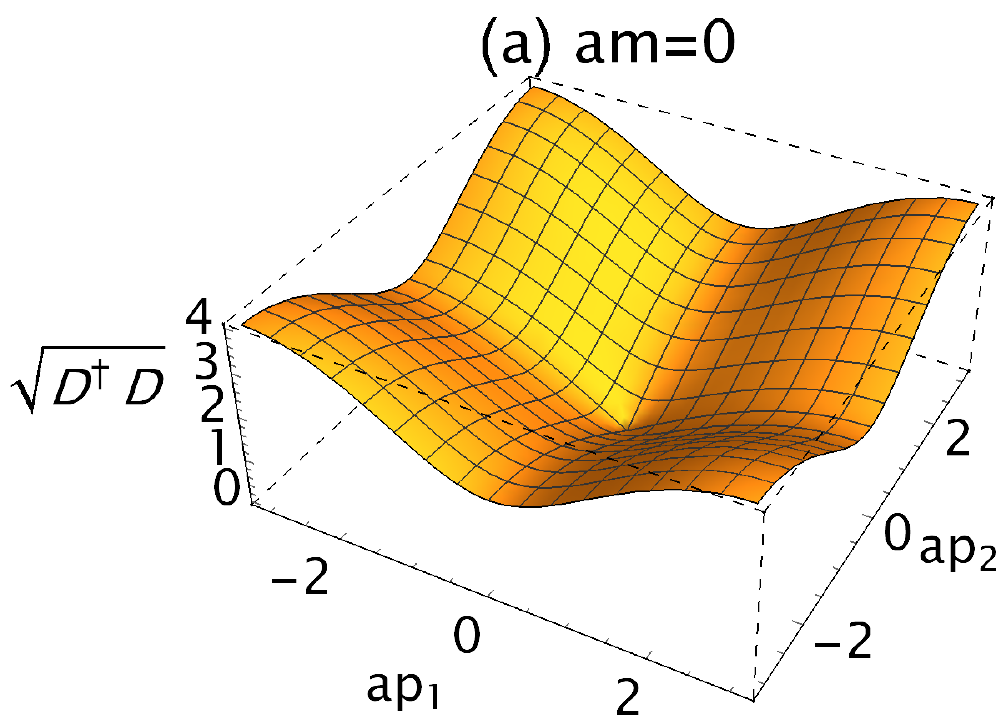}
            \includegraphics[clip, width=0.45\columnwidth]{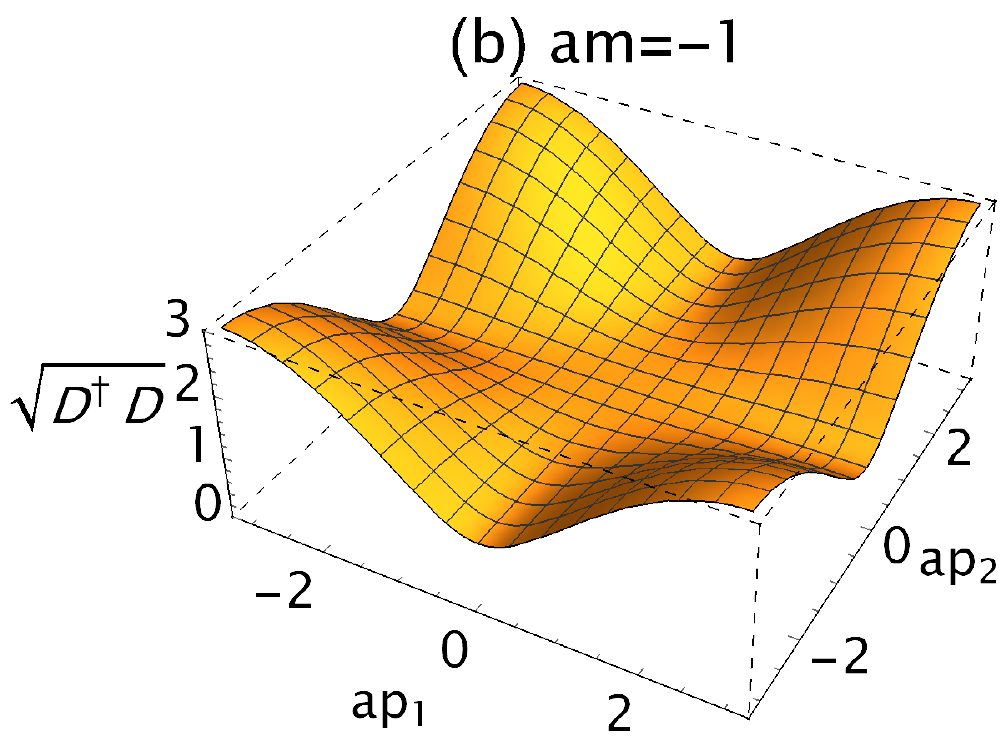}
        \end{center}
    \end{minipage}
    \begin{minipage}[t]{1.0\columnwidth}
        \begin{center}
            \includegraphics[clip, width=0.45\columnwidth]{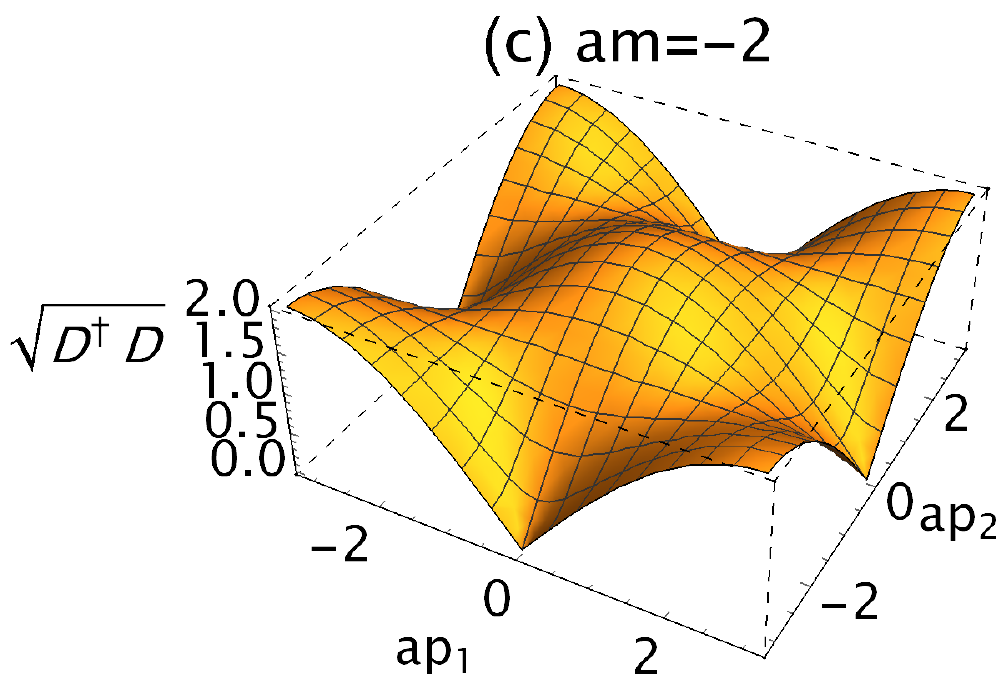}
            \includegraphics[clip, width=0.45\columnwidth]{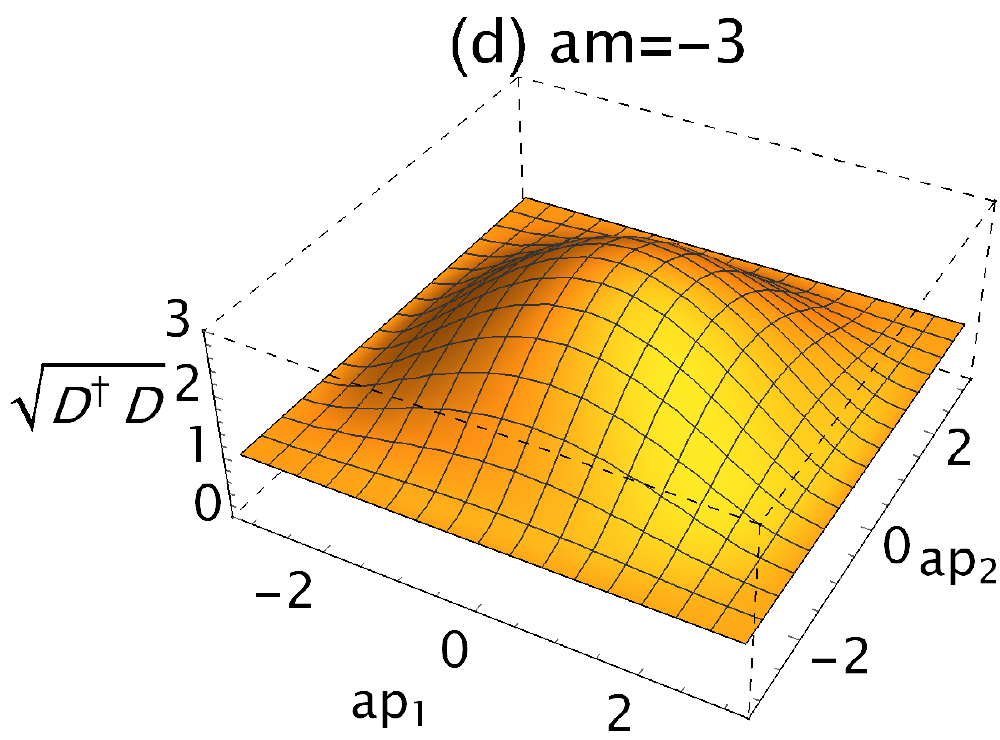}
        \end{center}
    \end{minipage}
    \begin{minipage}[t]{1.0\columnwidth}
        \begin{center}
            \includegraphics[clip, width=0.45\columnwidth]{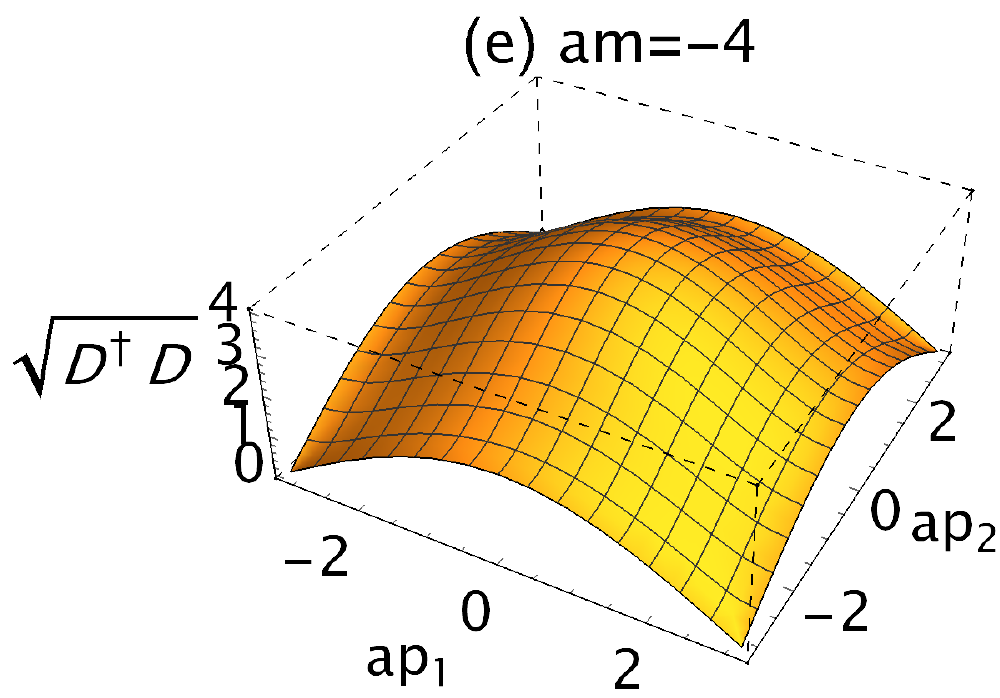}
            \includegraphics[clip, width=0.45\columnwidth]{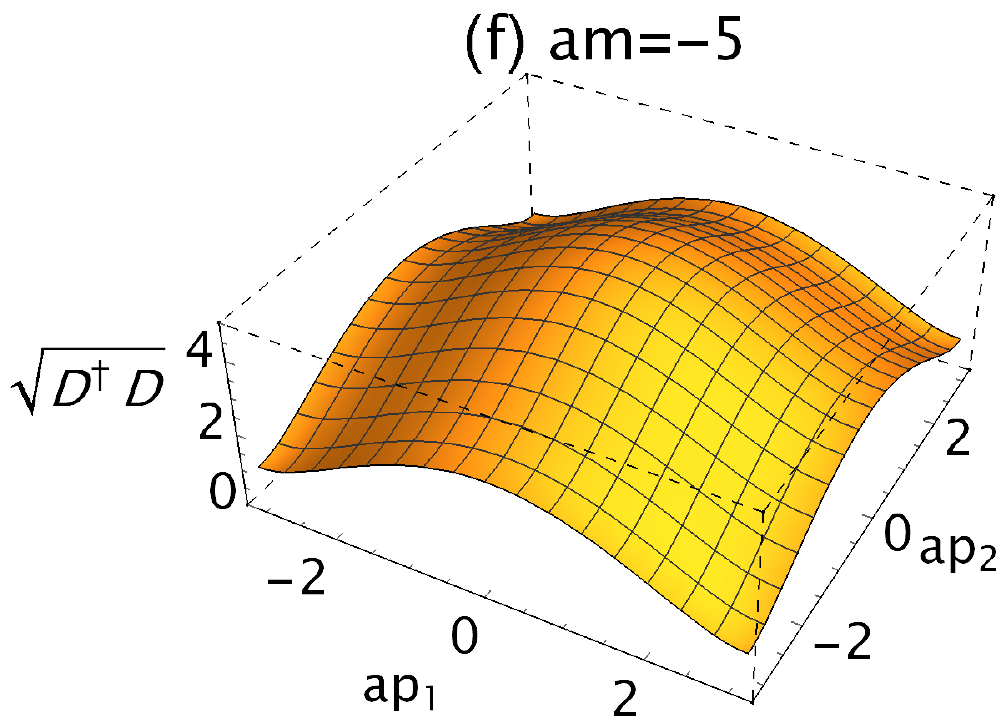}
        \end{center}
    \end{minipage}
    \caption{Negative-mass dependence of dispersion relations for Wilson fermions in the $2+1$-dimensional space-time (the temporal direction is not latticized).}
\label{fig:3d_Wil_neg_disp}
\end{figure}

For higher dimensions, the negative-mass dependence becomes more complicated.
In the lower panels in Fig.~\ref{fig:Wil_nega}, we show the numerical results in the $2+1$-dimensional space-time, where the spatial one dimension has the periodic or antiperiodic boundary condition, and another spatial dimension has no boundary.
Their corresponding dispersion relations are shown in Fig.~\ref{fig:3d_Wil_neg_disp}.
\begin{itemize}
\item[(1)]{\it $-2 < am_f \leq 0$}---At $am_f=0$, the Casimir energy $aE_\mathrm{Cas}$ is proportional to $1/N^2$ in larger lattice size, and its coefficient $NLE_\mathrm{Cas}$ is not suppressed.
Such a behavior is induced by the infrared zero mode, $(ap_1,ap_2)=(0,0)$, as shown in the left upper panel of Fig.~\ref{fig:3d_Wil_neg_disp}.
In the region of $-2 < am_f < 0$, $aE_\mathrm{Cas}$ is not proportional to $1/N^2$, and $NLE_\mathrm{Cas}$ is suppressed in larger lattice size.
This behavior implies that the relevant degrees of freedom may be massive.
For example, we show the dispersion with $am_f=-1$ in Fig.~\ref{fig:3d_Wil_neg_disp}.
From this figure, we see that the energies of all the modes are more than $1$, and there is no zero mode.
\item[(2)]{\it $am_f=-2$}---This is a special mass parameter.\footnote{The Wilson fermion satisfying $am_f+dr = 0$, where $d$ is the dimensions in Euclidean lattice space, is known as the {\it central-branch Wilson fermion}~\cite{Creutz:2011cd,Kimura:2011ik,Misumi:2012eh,Chowdhury:2013ux,Misumi:2019jrt,Misumi:2020eyx} which is a special lattice fermion.
In our setup, the time in the $2+1$-dimensional space-time is not on the lattice, so that the Wilson fermion with $am_f=-2$, studied in this section, is (partly similar to but) different from the central-branch Wilson fermion.}
Here, we find an oscillatory behavior of the Casimir energy between the odd and even lattices.
$NLE_\mathrm{Cas}$ is not suppressed in larger lattice size, which implies that the Casimir energy is dominated by massless degrees of freedom. 
As shown in left middle panel of Fig.~\ref{fig:3d_Wil_neg_disp}, there are two zero modes, $(ap_1,ap_2)=(0,\pi)$ and $(\pi,0)$.
Note that the other zero modes $(ap_1,ap_2)=(0,-\pi)$ and $(-\pi,0)$ are outside the first Brillouin zone ($-\pi< ap_k \leq \pi$), so that they do not contribute to the Casimir effect we showed.
\item[(3)]{\it $-4<am_f<-2$}---We find an oscillatory Casimir energy, where $NLE_\mathrm{Cas}$ is suppressed in larger lattice size.
We show $am_f=-3$ in the right middle panel of Fig.~\ref{fig:3d_Wil_neg_disp}, where we see that there is no zero mode, but there are lowest modes in the ultraviolet-momentum region ($|ap_k| \sim \pi)$.
These lowest modes contribute to the oscillation of the Casimir energy.
\item[(4)]{\it $am_f=-4$}---We also find an oscillatory behavior of the Casimir energy.
Here, $NLE_\mathrm{Cas}$ is not suppressed in larger lattice size, which is a similar behavior to $am_f=-2$.
As shown in the left lower panel of Fig.~\ref{fig:3d_Wil_neg_disp}, we can see that there are an ultraviolet-momentum zero mode, $(ap_1,ap_2)=(\pi,\pi)$.
Note that the other zero modes, $(ap_1,ap_2)=(\pi,-\pi)$, $(-\pi,\pi)$, and $(-\pi,-\pi)$, are outside the first Brillouin zone, so that they do not contribute to the Casimir effect.
\item[(5)]{\it $am_f<-4$}---In this region, we find an oscillatory Casimir energy.
$NLE_\mathrm{Cas}$ is suppressed in larger lattice size, which is similar to the case with $-4<am_f<-2$.
We show $am_f=-5$ in the right lower panel of Fig.~\ref{fig:3d_Wil_neg_disp}.
The form of the dispersion is similar to $am_f=-4$, but all the modes are more than $1$, and there is no zero mode.
\end{itemize}

Finally, we comment on the relationship to two-dimensional (2D) materials in condensed matter physics.
When we consider the $2\times 2$ component Wilson Dirac operator, $c_\mathrm{deg}=1$, the corresponding materials are quantum anomalous Hall insulators (or Chern insulators). 
These materials violate time-reversal invariance and can be described by the Qi-Wu-Zhang model~\cite{Qi:2005,Qi:2008} as a typical model.
Then, the Wilson fermion with either a positive mass $am_f>0$ or a large negative mass $am_f<-4$ is a normal insulator with trivial topology.
On the other hand, systems described by the Wilson fermion with $-2<am_f<0$ or $-4<am_f<-2$ correspond to quantum anomalous Hall insulators, where the nontrivial topology is characterized by TKNN integers~\cite{Thouless:1982zz} or (first) Chern numbers~\cite{Kohmoto:1985}, $C_1=1$ and $C_1=-1$, respectively.
Thus, $-2<am_f<0$ and $-4<am_f<-2$ are topological phases distinguished by a different Chern number.
The existence of an open boundary condition induces a gapless chiral edge mode along the one-dimensional edge of the insulator.
The Casimir effect for such an edge mode is not related to what we analyzed in this section, but it will be studied in Sec.~\ref{Sec_5}.
The Casimir effect for the 2D bulk modes corresponds to that studied in this section, which will be significant if the length of one direction of the 2D material is very short, and it can affect thermodynamic phenomena for the bulk modes.

As another example, when we consider the $4 \times 4$ component Wilson Dirac operator, $c_\mathrm{deg}=2$ (the case with Kramers doublet such as spin degrees of freedom), the corresponding materials are time-reversal invariant 2D topological insulators, namely quantum spin Hall insulators, which are two copies of quantum anomalous Hall insulators.
Such materials can be described by the Bernevig-Hughes-Zhang model~\cite{Bernevig:2006} and were experimentally observed in CdTe/HgTe/CdTe quantum well~\cite{Konig:2007}.
Even in this case, within our setup, the corresponding Casimir energy for the bulk fermion is qualitatively the same (except for the factor of the number of the degrees of freedom, $c_\mathrm{deg}=2$) as long as the dispersion relations of the different spin components degenerate.

\section{Casimir energy for overlap fermion} \label{Sec_5}
In this section, we study the Casimir energy for the overlap fermion with an MDW kernel operator.
In the DW fermion formulation~\cite{Kaplan:1992bt,Shamir:1993zy,Furman:1994ky}, a ``bulk" fermion is defined in the $D+1$-dimensional Euclidean space, which becomes a kernel operator in the $D$-dimensional space.
This bulk fermion is projected into the chiral ``surface" fermion in the $D$-dimensional space.
Usually, the surface fermions include information on the finite length of the extra dimension, but for simplicity we consider the infinite length.
Then the DW fermion is equivalent to the overlap fermion~\cite{Neuberger:1997fp,Neuberger:1998wv}.

\subsection{MDW kernel operators}
Here, we define the MDW kernel operator $D_\mathrm{MDW}$~\cite{Brower:2004xi,Brower:2005qw,Brower:2012vk},
\begin{align}
aD_\mathrm{MDW} \equiv \frac{b (aD_\mathrm{W})}{2 + c (aD_\mathrm{W})},
\end{align}
where $b$ and $c$ are called {\it M\"obius parameters}.
The operator at $b=1$ and $c=1$ corresponds to the conventional Shamir-type formulation~\cite{Shamir:1993zy}, and that at $b=2$ and $c=0$ is Bori\c{c}i-type (or Wilson-type)~\cite{Borici:1999zw,Borici:1999da}.
$D_\mathrm{W}$ is the Wilson Dirac operator with $r=1$, as defined in Eq.~(\ref{eq:Dw}), where the original fermion mass $m_f$ is replaced by the {\it domain-wall height} $am_f \to -M_0$ that plays a role as the negative mass of bulk fermions.

\subsection{Overlap fermion with MDW kernel}
Using the MDW kernel operator $D_\mathrm{MDW}$, we define the overlap Dirac operator $D_\mathrm{OV}$ with a fermion mass $m_f$ and a Pauli-Villars mass $m_\mathrm{PV}$,
\begin{align}
aD_\mathrm{OV}
\equiv&
(2 - cM_0)
M_0
am_\mathrm{PV} \nonumber\\
& \times \frac{(1 + am_f) + (1 - am_f)V }
{(1 + am_\mathrm{PV}) + (1 - am_\mathrm{PV})V},
\end{align}
with
\begin{align}
V \equiv \gamma_5\mathrm{sign}(\gamma_5 a D_\mathrm{MDW}) = \frac{D_\mathrm{MDW}}{\sqrt{D_\mathrm{MDW} ^\dagger D_\mathrm{MDW}}}. \label{Vsign}
\end{align}
The Pauli-Villars mass $m_\mathrm{PV}$ was introduced so as to satisfy the Ginsparg-Wilson relation.
The scaling factor $
(2 - cM_0) M_0 m_\mathrm{PV}$ with a constraint $2 - cM_0 > 0$ is determined so as to realize the dispersion relation of fermions in the continuum theory: $\lim_{a \rightarrow 0} D^\dagger _\mathrm{OV}D_\mathrm{OV} = p^2$ for $m_f = 0$.
Note that if we consider a finite length of the extra dimension, then the sign function in Eq.~(\ref{Vsign}) is replaced by an approximate functional form depending on the finite length.

As a result, the dispersion relation of the overlap fermion is written as
\begin{align}
a^2E^2_\mathrm{OV}
 =&
\left[(2 - cM_0)
M_0 m_\mathrm{PV} \right]^2 \nonumber\\
& \times \frac{2[1 + (am_f)^2] + [1 - (am_f)^2] (V^\dagger + V)}
{2[1 + m_\mathrm{PV}^2] + [1 - m_\mathrm{PV}^2](V^\dagger + V)}. \label{eq:dispOV}
\end{align}
where we used $V^\dagger V = 1$ and the commutation relation between $V^\dagger + V $ and $V$.
Using $D_\mathrm{W}$, $V^\dagger + V$ is
\begin{align}
V^\dagger + V
=&
\left(D_\mathrm{MDW} ^\dagger + D_\mathrm{MDW} \right)\frac{1}{\sqrt{D_\mathrm{MDW} ^\dagger D_\mathrm{MDW}}}, \notag\\
=&
2(D_\mathrm{W} + D_\mathrm{W} ^\dagger + cD_\mathrm{W}^\dagger D_\mathrm{W})
\frac{1}
{\sqrt{D_\mathrm{W} ^\dagger D_\mathrm{W}}} \nonumber\\
&\times
\frac{1}{\sqrt{{4 + 2c(D_\mathrm{W} ^\dagger + D_\mathrm{W}) + c^2 D_\mathrm{W} ^\dagger D_\mathrm{W}}}}, \label{eq:Vdag+V}
\end{align}
where we used $D_\mathrm{W}^\dagger D_\mathrm{W} > 0$ and ${4 + 2c(D_\mathrm{W} ^\dagger + D_\mathrm{W}) + c^2 D_\mathrm{W} ^\dagger D_\m{W}} > 0$.

Note that, in this work, the overlap Dirac operator is independent of the parameter $b$.
The $b$ dependence can appear when we consider a finite length of the extra dimension.
By substituting the square root of the dispersion relation (\ref{eq:dispOV}) into the definitions, (\ref{eq:def_cas}) and (\ref{eq:def_cas1+1}), we can calculate the Casimir energy for the overlap fermion.

\subsection{Numerical results}
\begin{figure*}[t!]
    \begin{minipage}[t]{1.0\columnwidth}
        \begin{center}
            \includegraphics[clip, width=1.0\columnwidth]{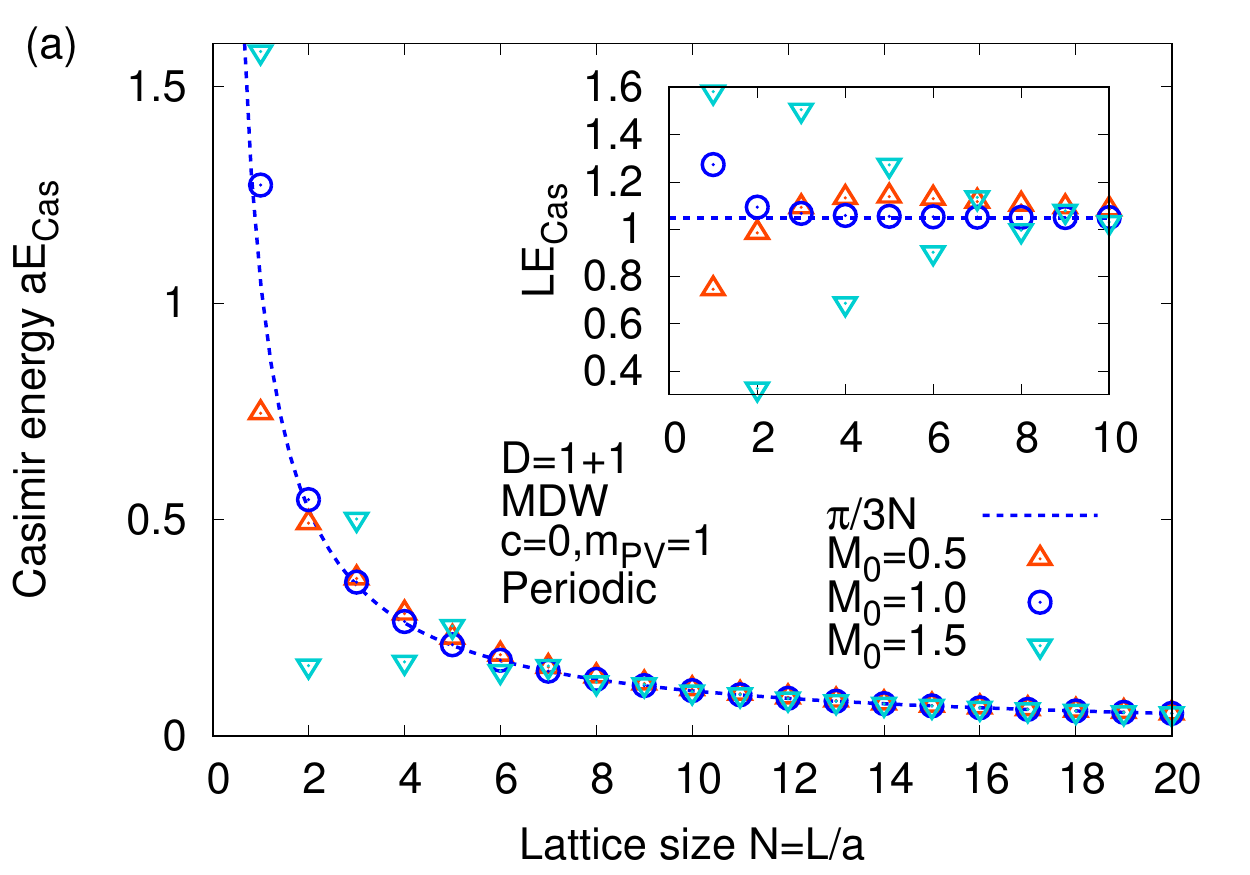}
            \includegraphics[clip, width=1.0\columnwidth]{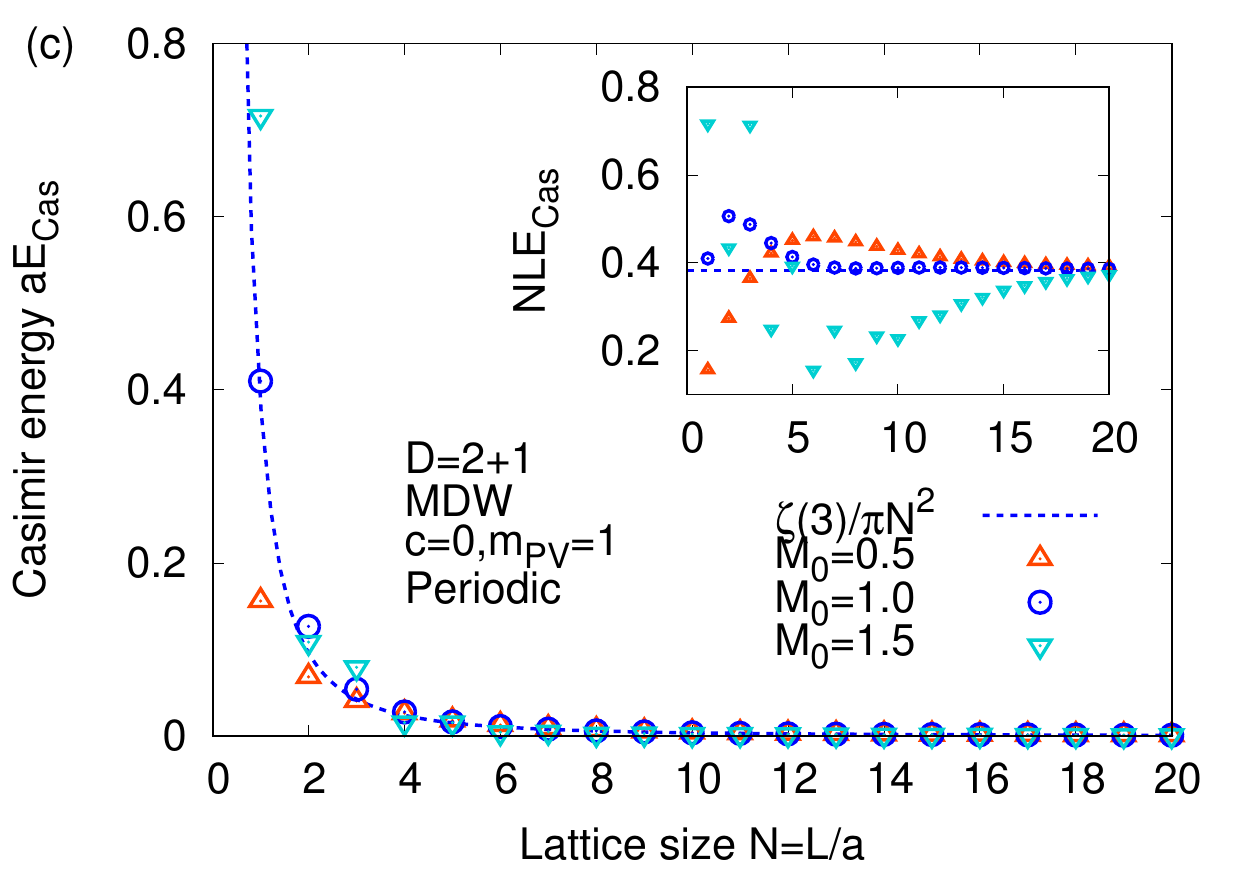}
            \includegraphics[clip, width=1.0\columnwidth]{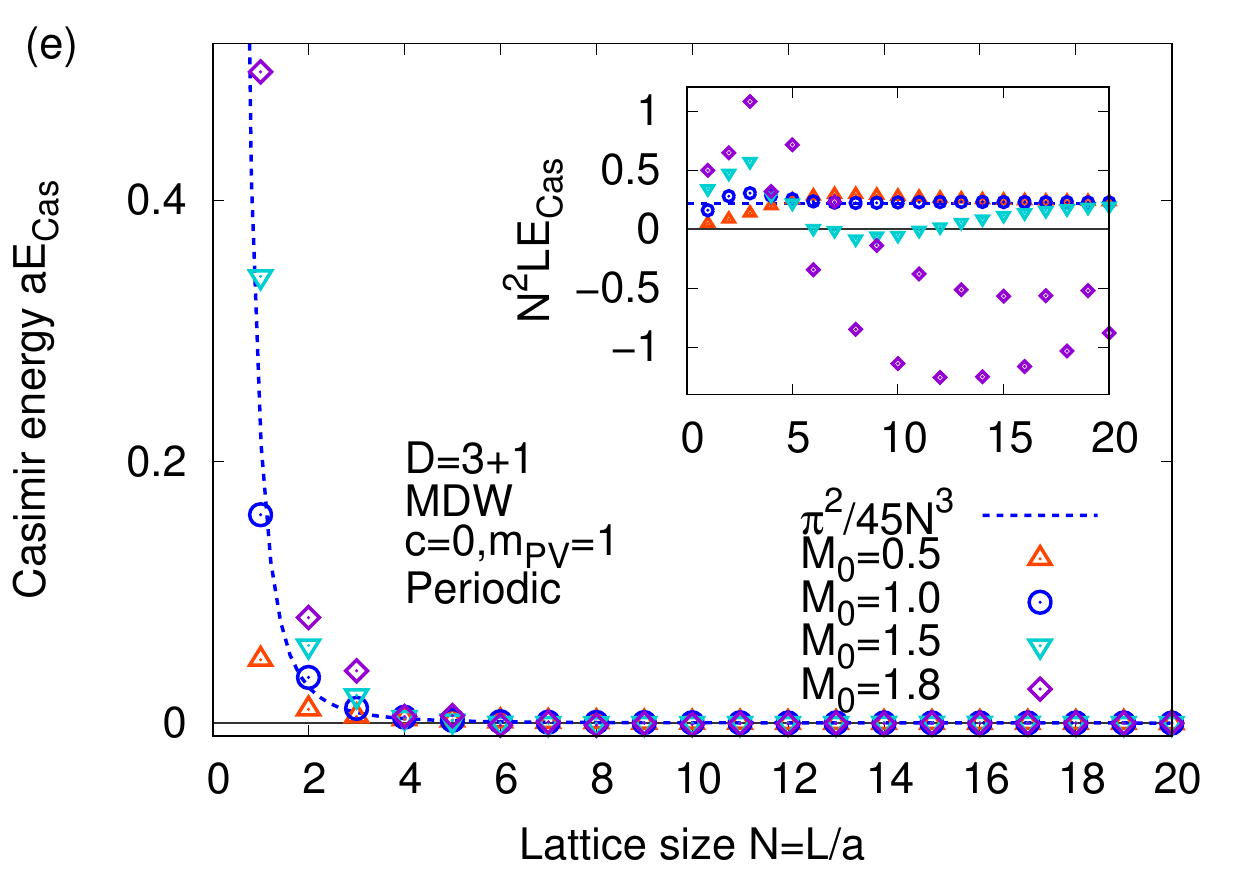}
        \end{center}
    \end{minipage}
    \begin{minipage}[t]{1.0\columnwidth}
        \begin{center}
            \includegraphics[clip, width=1.0\columnwidth]{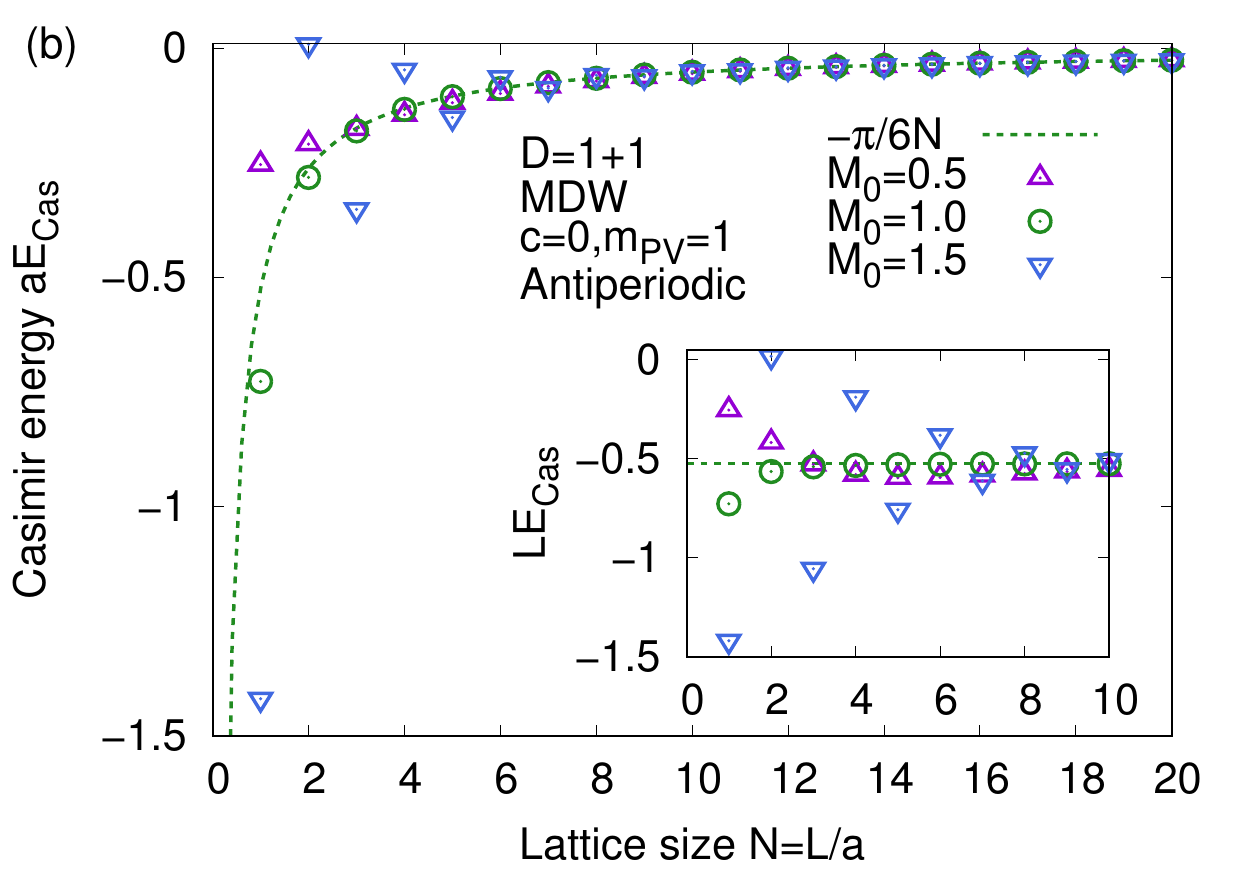}  
            \includegraphics[clip, width=1.0\columnwidth]{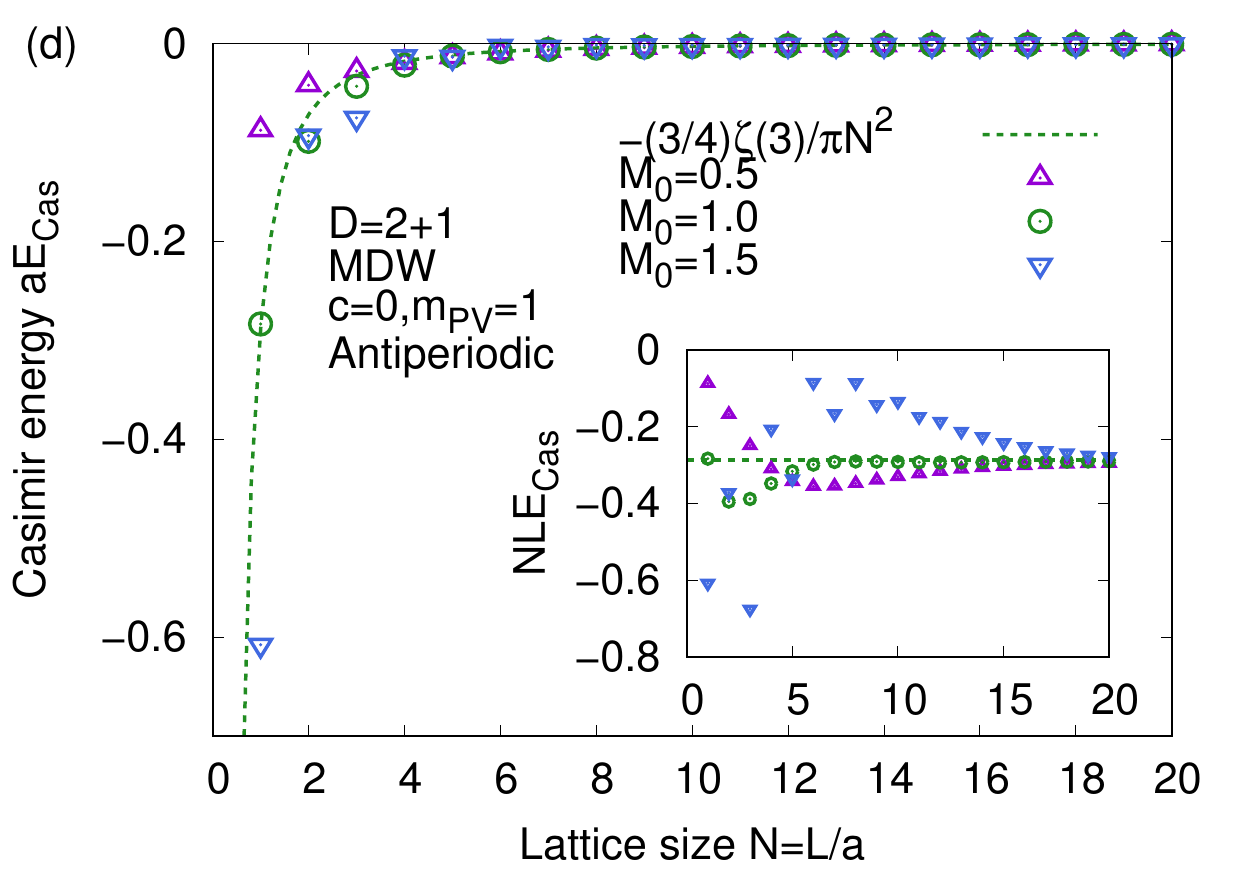}
            \includegraphics[clip, width=1.0\columnwidth]{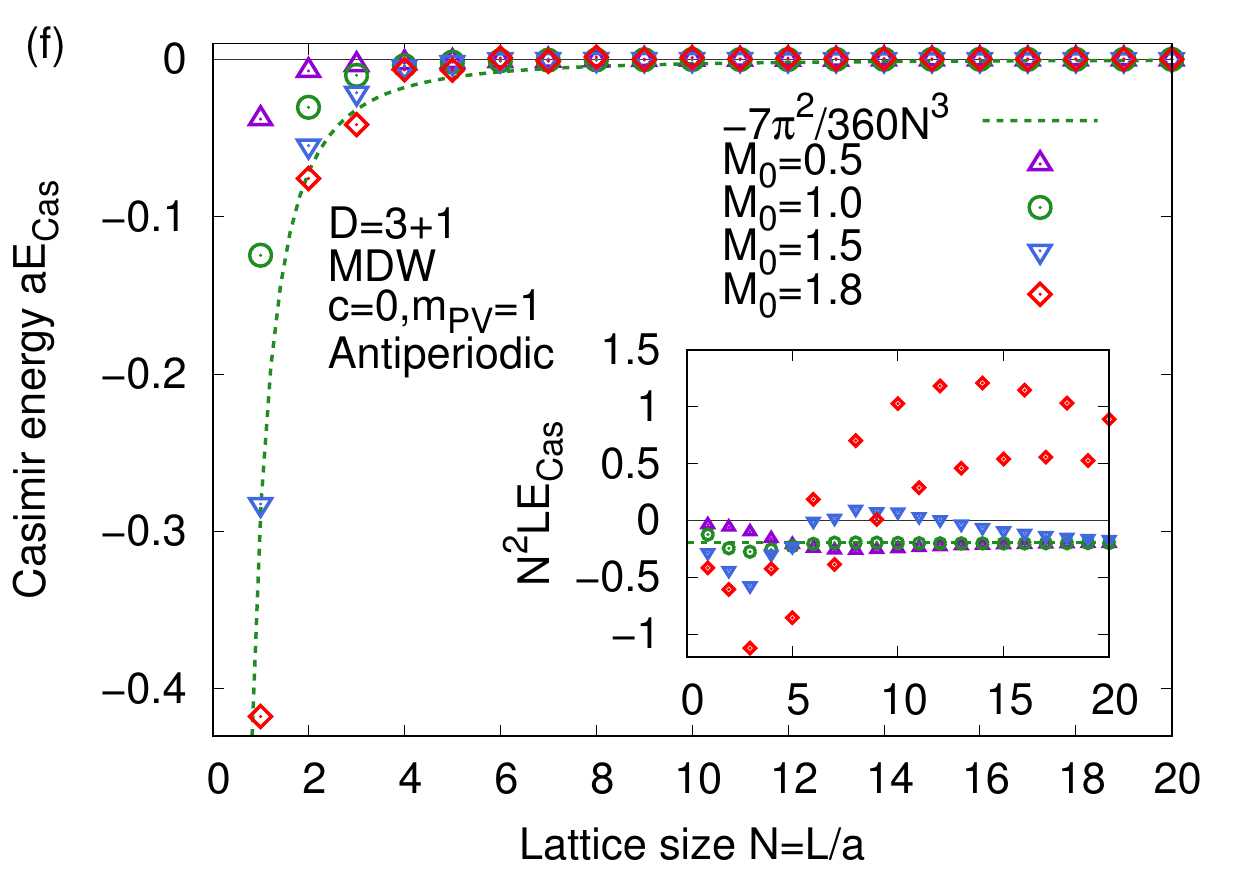}    
        \end{center}
    \end{minipage}
    \caption{Casimir energy for overlap fermions with MDW kernel operator in the $1+1$-, $2+1$-, and $3+1$-dimensional space-time (the temporal direction is not latticized).
M\"obius parameter $c=0$ and Pauli-Villars mass $m_\mathrm{PV}=1$ are fixed.
Small windows show the coefficients of Casimir energy.
Dashed lines are the leading terms of the expansion by $a/L$ or equivalently $1/N$, which is obtained as an asymptotic form for the massless fermion in the large lattice size $N$.
(Left) Periodic boundary. (Right) Antiperiodic boundary.
}
\label{fig:2d3d4d_MDWm0}
\end{figure*}

\begin{figure}[t!]
    \begin{minipage}[t]{1.0\columnwidth}
        \begin{center}
            \includegraphics[clip, width=1.0\columnwidth]{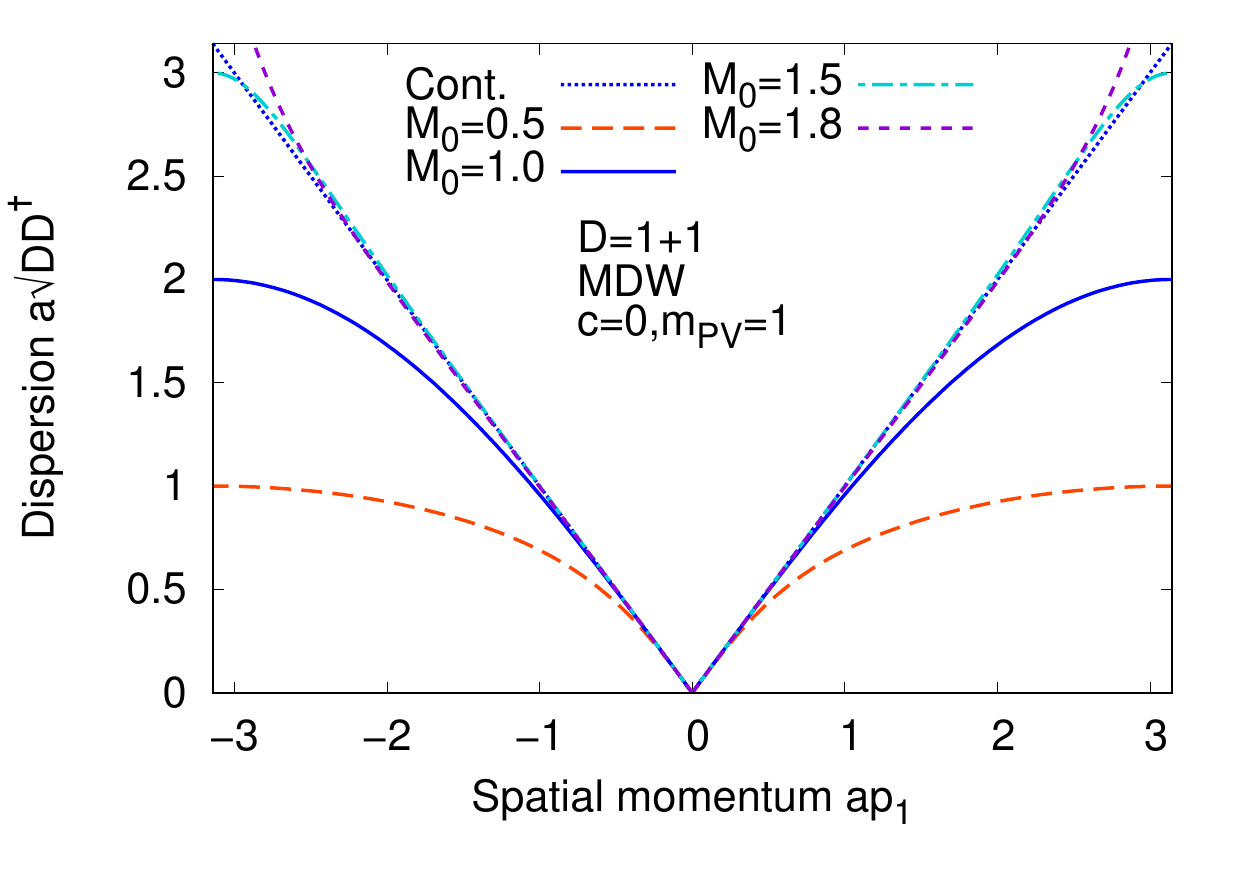}
        \end{center}
    \end{minipage}
    \caption{Domain-wall height $M_0$ dependence of dispersion relations for overlap fermions with MDW kernel operator in the $1+1$-dimensional space-time (the temporal direction is not latticized).}
\label{fig:2d_MDWm0_disp}
\end{figure}

In Fig.~\ref{fig:2d3d4d_MDWm0}, we show the dependence of the Casimir energy for the overlap fermion with the MDW kernel operator on the domain-wall height ($M_0 =0.5$, $1.0$, and $1.5$), where $c=0$ and $m_\mathrm{PV}=1.0$ are fixed.
The results in the $1+1$ dimensions are the same as that obtained in Ref.~\cite{Ishikawa:2020ezm}, and the results in the $2+1$ or $3+1$ dimensions are newly obtained in this work.
In Fig.~\ref{fig:2d_MDWm0_disp}, we show the corresponding dispersion relations in the $1+1$-dimensional space-time.
From this figure, we see that the ultraviolet part of the dispersion relations is modified by tuning $M_0$.
Note that $M_0 =1.0$ in the $1+1$ dimensions is equivalent to the Wilson fermion at $am_f=0$.

First, we summarize the properties of the Casimir energy in the three regions (namely, small $N$, intermediate $N$, and large $N$):
\begin{itemize}
\item[(1)]{\it Suppression of Casimir energy in small $N$}---In Fig.~\ref{fig:2d3d4d_MDWm0}, for $M_0 \lesssim 1.0$, we find that the Casimir energy for the overlap fermion is suppressed as the lattice size $N$ decreases.
Such suppression of the Casimir energy at a small lattice size is intuitively understood by considering the case at $N=1$.

For example, for the periodic boundary in the $1+1$ dimensions, the possible momentum and energy at $N=1$ are only $ap_1=0$ and $aE=0$, respectively.
Then the sum part of $aE_\mathrm{Cas}$ defined as Eq.~(\ref{eq:def_cas1+1}) is zero, so that $aE_\mathrm{Cas}$ is determined by only the positive integral part.
The integral part at a smaller $M_0$ is smaller than that at a larger $M_0$, so that the corresponding (positive) Casimir energy at the smaller $M_0$ is also suppressed compared with the larger $M_0$.

The case with the antiperiodic boundary is more complicated than that with the periodic boundary.
In the $1+1$ dimensions, the possible momentum at $N=1$ is only $ap_1=\pi$ which is the maximum energy level within the Brillouin zone, and this energy level dominates the negative sum part of $aE_\mathrm{Cas}$.
The difference between the negative sum part and positive integral part determines the negative $aE_\mathrm{Cas}$.
With decreasing $M_0$, both the parts are suppressed, where the negative $aE_\mathrm{Cas}$ is determined by a balance between the two parts.
As a result, $aE_\mathrm{Cas}$ is also smaller.
Thus a small $M_0$ can lead to the suppression of the Casimir energy.

\item[(2)]{\it Enhancement of Casimir energy in intermediate $N$}---Next we focus on the intermediate $N$ region.
For $M_0 \lesssim 1.0$, we find that the Casimir energy for the overlap fermion is enhanced, compared with that expected in the continuum theory.
Such enhancement is similar to that for the Wilson fermion at $am_f=0$ in the small $N$ region.
For the periodic boundary, this enhancement is induced by contributions from nonzero modes ($ap_1>0$) which appears at $N \geq 2$.
For the antiperiodic boundary, this is caused by contributions from lower modes ($ap_1<\pi$) except for the maximum mode ($ap_1=\pi$).

\item[(3)]{\it Good agreement with continuum theory in large $N$}---In the large $N$ region, the Casimir energy for the overlap fermion agrees well with that in the continuum theory, which is similar to the Wilson fermion at $am_f=0$.
Thus the Casimir energy in the large $N$ region is determined by the infrared part of the dispersion relations.
\end{itemize}

Next, we summarize the domain-wall height $M_0$ dependence:
\begin{itemize}
\item[(1)]{\it No oscillation at $M_0<1.0$}---First, at $M_0=0$, the overlap fermion is not defined.
In the region with $0< M_0 \leq 1.0$, we find that there is no oscillation.
This is because the doublers are sufficiently massive, so that their contributions are irrelevant.
When $M_0 \neq 0$ is small enough, both the suppression of Casimir energy at small $N$ and the enhancement at intermediate $N$ are visible.
\item[(2)]{\it Good agreement with continuum theory at $M_0=1.0$}---In the $1+1$-dimensional space-time, the Casimir energy at $M_0=1.0$ is equivalent to that for the Wilson fermion and well reproduces that in the continuum theory.
On the other hand, in the higher dimensions, the Casimir energy for the overlap fermion is not equivalent to that for the Wilson fermion, but they also well reproduce that in the continuum theory.
\item[(3)]{\it Oscillation at $M_0>1.0$}---In the region with $M_0>1.0$, we find an oscillatory behavior of the Casimir energy, where we find the enhancement on the odd lattice and suppression on the even lattice.
This oscillation is induced by the contributions from massive doublers\footnote{In the domain-wall fermion formulation, the doublers with a heavy mass still survive.
We call such ultraviolet-momentum modes ``massive doublers".} which is irrelevant at $M_0 \leq 1.0$.
As shown in Fig.~\ref{fig:2d_MDWm0_disp}, the dispersion relation at $M_0 = 1.5$ is very close to the linear dispersion, but the corresponding Casimir energy in the small $N$ region is quite different.
Therefore, if we try to reproduce the Casimir energy for the continuum Dirac fermion by using the overlap fermion, $M_0 \sim 1.0$ is more suitable than larger $M_0$.

In lattice simulations for fermions interacting with gauge fields, a setup with $M_0>1.0$ may be useful (e.g., see Refs.~\cite{Blum:2000kn,Aoki:2002vt,Brower:2004xi,Brower:2005qw,Brower:2012vk}).
This is because a setup with $M_0>1.0$ can correct the fermion dispersion relations effectively modified by gauge fields.

Note that a similar oscillation for the overlap fermion with the domain-wall kernel operator was also seen in free fermions at finite temperature~\cite{Banerjee:2008ii,Gavai:2008ea} which corresponds to the antiperiodic boundary for the temporal direction in the Euclidean lattice.
\end{itemize}

Finally, we comment on the dependence on the M\"obius parameter $c$ and the Pauli-Villars mass $m_\mathrm{PV}$ as additional tuning parameters.
The results in the $1+1$ dimensions are shown in Ref.~\cite{Ishikawa:2020ezm}.
While the M\"obius parameter $c$ does not induce oscillation of the Casimir energy, a large value of the Pauli-Villars mass $m_\mathrm{PV}$ leads to an oscillation of the Casimir energy, which is a similar mechanism to the massive doublers in $M_0>1.0$.

\section{Conclusion and outlook} \label{Sec_6}
In this paper, based on the formulation given by Ref.~\cite{Ishikawa:2020ezm}, we investigated the Casimir energies for the free lattice fermions in the $1+1$-, $2+1$-, and $3+1$-dimensional space-time with the one spatial direction compactified by the periodic or antiperiodic boundary condition.
In particular, the Casimir energy for the Wilson fermion with a negative mass was examined for the first time.
The results in the $2+1$, and $3+1$ dimensions are almost similar to those in the $1+1$ dimensions~\cite{Ishikawa:2020ezm}, but we also found a special property realized only in the $1+1$ dimensions, such as Eq.~(\ref{eq:ECas_am=-1}).

We again emphasize that the Wilson fermion with a negative mass and the domain-wall fermion correspond to the bulk and surface fermions of topological insulators, respectively.
Therefore the Casimir effects for their lattice fermions will be observed as the Casimir effects on the topological insulators if the very small lattice size, such as thin films, is experimentally realized.
Then the oscillatory behavior of Casimir energy, which we suggested, may be helpful to capture an evidence of the Casimir effect on the lattice.
Similar oscillations were suggested also between localized impurities in one-dimensional materials~\cite{Fuchs:2007,Wachter:2007,Kolomeisky:2008,Zhabinskaya:2009}.

As examples of experimental observables for the Caimir effect, we can consider an influence on the specific heat and magnetic susceptibility of materials.
The theoretical strategy is as follows.
In a standard manner, we can calculate the thermodynamic potential of a lattice fermionic system.
When we switch on the finite-volume and finite-temperature effect, the resultant potential depends on the temperature and system size, and its system size dependence is related to the Casimir effect, in principle.
Since the specific heat is the temperature-derivative of the potential, we can estimate the change of the specific heat by the Casimir effect.
Similarly, from the second derivative of the potential with respect to the external magnetic field, we can calculate the magnetic susceptibility.
Thus, the Casimir effect, in principle, can contribute to various physical quantities, and we can observe it if the system size is small enough.

Here, we comment on the realistic scale of the Casimir energy.
The Casimir energy $aE_\mathrm{Cas}$ in this paper is a dimensionless quantity, and as a dimensional quantity, we consider $E_\mathrm{Cas} \sim \hbar c/aN^d$ in elemental particles or $E_\mathrm{Cas} \sim \hbar v_F/aN^d$ in solid states, where $\hbar$, $c$, and $v_F$ are the reduced Plank constant, the speed of light, and the Fermi velocity, respectively, and we omitted the specific coefficient depending on the types of fermions and spatial dimensions as $\mathcal{O}(1)$.
For example, for Cd${}_{3}$As${}_{2}$ known as a three-dimensional Dirac semimetal~\cite{Wang:2013,Neupane:2014}, the lattice spacing is $a=1$-$3$ nm, and the Fermi velocity was measured to be $v_F \sim 1.5 \times 10^6$ m/s~\cite{Neupane:2014}.
Using these values, the Casimir energy for Cd${}_{3}$As${}_{2}$ is estimated to be $E_\mathrm{Cas} \sim 1$ eV at $N=1$.
Therefore, it will be captured by an external response comparable to this scale.
As numerical values of other materials, $a=0.5$-$1$ nm and $v_F \sim (1$-$5) \times 10^5$ m/s~\cite{Liu:2014} for Na${}_{3}$Bi, and $a=0.4$-$3$ nm and $v_F \sim 5 \times 10^5$ m/s (for the surface modes)~\cite{Xia:2009zza} for Bi${}_{2}$Se${}_{3}$.

One of the extensions is to study modification of Casimir effects by introducing an interaction between fermions, such as gauge interactions and four-Fermi interactions.
In particular, interacting Wilson fermions with a negative mass exhibit the spontaneous parity-broken (Aoki) phase~\cite{Aoki:1983qi}.
Since the negative-mass Casimir effect studied in this paper is just for free fermions, its phase structure switching on an interaction would be interesting.

Furthermore, cold-atom simulations can realize various lattice fermions \cite{Bermudez:2010da,Mazza:2011kf,Kuno:2018rmc,Zache:2018jbt}, so that simulations with a small size will be a powerful tool to observe the Casimir effect on the lattice. 

In this work, we focused on only 1D chain, 2D square, or 3D cubic lattices, but other spatial geometries are possible.
For example, lattice structures such as (2D) honeycomb or (3D) diamond lattices are also interesting.
Casimir effects on honeycomb lattices such as graphene nanoribbons,\footnote{For the {\it electromagnetic} Casimir effects induced between graphene and another material (or between two graphene sheets), there are already many works (e.g., see Refs.~\cite{Bordag:2009fz,Gomez-Santos:2009rxl,Drosdoff:2010avo,Fialkovsky:2011pu,Sernelius:2011zz}).} and carbon nanotubes,\footnote{Carbon nanotubes {\it in the continuum limit} can be approximated as the cylindrically compactified space with the spatial topology $R^1 \times S^1$ or the toroidally compactified space with the spatial topology $S^1 \times S^1$.
The Casimir effect for Dirac fermions in such continuous space-time was investigated in Refs.~\cite{Bellucci:2009jr,Bellucci:2009hh,Bellucci:2010xd,Elizalde:2011cy,Bellucci:2012za,Bellucci:2013oya}.
If one tries to investigate more realistic carbon nanotubes, systems with nonzero lattice spacing should be taken into account.} quantum anomalous Hall insulators described by the Haldane model~\cite{Haldane:1988zza}, and quantum spin Hall insulators described by the Kane-Mele model~\cite{Kane:2004bvs,Kane:2005zz} will be experimentally measured in small-size materials.
In particular, when we consider the edge of the small-size direction of honeycomb lattice ribbons, there are two types of structures: the armchair edge and the zigzag edge.
These two types of edges are known to lead to different band structures of bulk modes~\cite{Fujita:1996}, and the size dependence of observables is related to the Casimir effect for the bulk modes.
Moreover, a rolled-up graphene sheet is a carbon nanotube, which is nothing but a 2D honeycomb lattice with the periodic boundary condition, and electrons living on this lattice are affected by the Casimir effect with the periodic boundary.
For carbon nanotubes, there are three types of structures: the armchair, zigzag, and chiral configurations are known to exhibit different band gap energies~\cite{Hamada:1992,Saito:1992}.
As a result, in a small-size tube, such three structures lead to different Casimir effects, and then it should influence transport/thermodynamic properties such as electric/thermal conductivity and specific heat.
In addition, Casimir effects on ``curved" lattice deformed by insertion of structural defects such as lattice kirigami~\cite{Castro:2018iqt,Flachi:2019btk} are also interesting, which is analogous to the Casimir effect in curved space-time in continuum theory. 
Thus a wide range of lattice Casimir physics is left for future work.

\section*{Acknowledgments}
The authors are grateful to Yasufumi Araki for carefully reading our manuscript and Daiki Suenaga for giving us helpful comments about the Abel-Plana formulas.
This work was supported by Japan Society for the Promotion of Science (JSPS) KAKENHI (Grant Nos. JP17K14277 and JP20K14476).

\appendix
\begin{widetext}
\section{Derivation of Casimir energy for massless fermions in continuum limit} \label{App:1}
We quickly review the derivation of the Casimir energy for a massless fermion in $D=d+1$-dimensional space-time.
The dispersion relation of the fermion is
\begin{align}
    E = \sqrt{p^2},
\end{align}
where $p^2=p_1^2 +p_2^2 + \cdots + p_d^2$.
The thermodynamic potential $\Omega$ at zero temperature and zero chemical potential is the integral of the dispersion relation:
\begin{align}
    \frac{\Omega}{V} = -\int\frac{d^d p}{(2\pi)^d} \sqrt{p^2},
\end{align}
where $V$ is the volume of the system.
It can be considered as the sum of the zero-point energies for all momenta.
The coefficient $1/2$ of the zero-point energy is canceled by the factor $2$ of the particle and antiparticle.
The negative sign is caused by the fermionic property. 
Notice that, for the Dirac fermion in the $3+1$ dimensions, the spin degeneracy factor $c_{\text{deg}}=2$ has to be multiplied, but we set $c_{\text{deg}}=1$ and neglect the spin degrees of freedom for simplicity.
The thermodynamic potential is still divergent, and usually one neglects it.
In the compactified space-time, however, we can get a finite value by a subtraction scheme, and it modifies thermodynamic properties. 

Let us consider the spatial geometry where one spatial dimension, $x$, is compactified.
Then the boundary condition is necessary.
Here we limit ourselves to the periodic boundary condition (PBC) and the antiperiodic boundary condition (ABC).
The momentum along the $x$ direction is discretized and depends on the boundary condition:
\begin{align}
    p_1 &= \frac{2\pi}{L}n \ (\text{PBC}),\\
    p_1 &= \frac{2\pi}{L}\left(n+\frac{1}{2} \right) \ (\text{ABC}),
\end{align}
where $L$ is the length of the $x$ direction, and $n$ is an integer.
We replace the integral by the sum:
\begin{align}
    \int\frac{d p_1}{2\pi}\to \frac{1}{L}\sum_{n=-\infty}^{\infty}.
\end{align}
The momenta along the other direction are continuous and integral variables.
We can compute the thermodynamic potential in the anisotropic system, the so-called Casimir energy.

The derivation of the Casimir energy is based on the analyticity of the dispersion relation.
The potential density
\begin{align}
    \label{eq:Omega_L}
    \frac{\Omega(L)}{V} = -\frac{1}{L}\sum_{n=-\infty}^{\infty}\int\frac{d^{d-1} p_\perp}{(2\pi)^{d-1}} \sqrt{p_\perp^2+p_1^2}
\end{align}
is still divergent, where $p_\perp^2= p_2^2 + \cdots +p_d^2$.
We have to extract a finite part by removing the divergence in the infinite volume from Eq.~\eqref{eq:Omega_L}.
In the massless case, it can be done by the zeta function regularization and the analytic continuation.

Let us move on to the derivation of the Casimir energy.
By using the well-known formula for a $d$ dimensional integral with respect to $k$, for parameters $l$ and $\Delta$,
\begin{align}
    \int\frac{d^{d} k}{(2\pi)^{d}} \left( k^2 +\Delta \right)^{-l}
    =\frac{1}{(4\pi)^{d/2}}\frac{\Gamma(l-\frac{d}{2})}{\Gamma(l)}
    \Delta^{\frac{d}{2}-l},
\end{align}
and with the replacement $\sqrt{p_\perp^2+p_1^2}\to \left( p_\perp^2+p_1^2 \right)^{-s}$ with a new parameter $s$, we can integrate the potential density:
\begin{align}
\frac{\Omega(L;s)}{V} &= -\frac{1}{L}\sum_{n=-\infty}^{\infty}\int\frac{d^{d-1} p_\perp}{(2\pi)^{d-1}} \left(p_\perp^2+p_1^2 \right)^{-s}\\
&= -\frac{1}{L}\sum_{n=-\infty}^{\infty}\frac{1}{(4\pi)^{(d-1)/2}}\frac{\Gamma(s-\frac{d-1}{2})}{\Gamma(s)} |p_1|^{d-1-2s}.
\end{align}
For the PBC, the sum of $p_1$ can be replaced by the zeta function through the analytic continuation:
\begin{align}
    \frac{1}{L}\sum_{n=-\infty}^{\infty}|p_1|^{d-1-2s}=\frac{2}{L}\sum_{n=1}^{\infty} \left(\frac{2\pi}{L}n \right)^{d-1-2s}=\frac{2}{L} \left( \frac{2\pi}{L} \right)^{d-1-2s}\zeta(2s-d+1).
\end{align}
Then the potential can be the simpler form
\begin{align}
    \label{eq:before_refrection}
    \frac{\Omega(L;s)}{V} 
    &= -\frac{2}{L}\frac{1}{(4\pi)^{(d-1)/2}}\frac{\Gamma(s-\frac{d-1}{2})}{\Gamma(s)}
    \left( \frac{2\pi}{L} \right)^{d-1-2s}\zeta(2s-d+1) \\
    \label{eq:after_refrection}
    &=-\frac{2}{L}\frac{1}{(4\pi)^{(d-1)/2}}\frac{\pi^{-1/2}\Gamma\left( \frac{d-2s}{2} \right) \zeta(d-2s)}{\Gamma(s)}
    \frac{2^{d-1-2s}}{L^{d-1-2s}},
\end{align}
where, in order to avoid the divergence of $\Gamma(-1)$ at $d=2$ in the limit $s \to -1/2$, we have applied the reflection formula
\begin{align}
    \Gamma(z/2)\zeta(z)=\pi^{z-1/2}\Gamma\left( \frac{1-z}{2} \right) \zeta(1-z).
\end{align}
Eq.~(\ref{eq:after_refrection}) is already finite even in the limit $s \to -1/2$, and the final form is
\begin{align}
    \frac{\Omega(L)}{V} &=\frac{2}{(L\sqrt{\pi})^{d+1}}\Gamma \left( \frac{d+1}{2} \right) \zeta(d+1).
\end{align}
We show the results in $d=1, 2, 3$:
\begin{align}
    \frac{\Omega(L)}{V} &=\frac{\pi}{3L^2}\ \ \ (d=1),\\
    \frac{\Omega(L)}{V} &=\frac{\zeta(3)}{\pi L^3} \ \ \ (d=2),\\
    \frac{\Omega(L)}{V} &=\frac{\pi^2}{45 L^4}\ \ \ (d=3).
\end{align}
By multiplying these formulas by $L$, we can obtain $E_\mathrm{Cas}$ discussed in the main text.

For the ABC, we just have to replace the zeta function $\zeta(2s-d+1) $ by the Hurwitz zeta function $\zeta(2s-d+1,1/2) $ in Eq.~(\ref{eq:before_refrection}).
This function can be rewritten in terms of the Riemann zeta function through $\zeta(z,1/2)=(2^z-1)\zeta(z)$.
Then, we can take the limit $s \to -1/2$ in the same way.
Finally, the resulting expression is
\begin{align}
    \frac{\Omega(L)}{V} &=-\frac{2^d-1}{2^d}\frac{2}{(L\sqrt{\pi})^{d+1}}\Gamma\left( \frac{d+1}{2} \right) \zeta(d+1).
\end{align}
We also show the expressions in $d= 1, 2, 3$:
\begin{align}
    \frac{\Omega(L)}{V} &=-\frac{\pi}{6L^2}\ \ \ (d=1),\\
    \frac{\Omega(L)}{V} &=-\frac{3\zeta(3)}{4\pi L^3} \ \ \ (d=2),\\
    \frac{\Omega(L)}{V} &=-\frac{7\pi^2}{360 L^4} \ \ \ (d=3).
\end{align}

\section{Abel-Plana formulas in finite range} \label{App:2}
The Abel-Plana formula (APF) is a conventional and powerful tool to study the Casimir effect (for early works, see Refs.~\cite{Mamaev:1976zb,Mamaev:1979ks,Mamaev:1979um,Mamaev:1979zw,Mamaev:1980jn}).
Usually, this formula is used to calculate the finite Casimir energy from an infinite integral and an infinite sum.
For lattice fermions, the momentum of a fermion has a periodicity within a Brillouin zone, and then the momentum space can be restricted to the first Brillouin zone. 
Therefore the APF should also be modified as that {\it in a finite range}.
In this appendix, we derive the APF in a finite range.
Our derivation is based on the following formula (see, Refs.~\cite{Saharian:2000xx,Saharian:2007ph}):
\begin{align}
  \label{eq:GAPF}
  \int_{a}^{b} f(x) d x=R[f(z), g(z)]-\frac{1}{2} \int_{-i \infty}^{+i \infty}[g(u)+\sigma(z) f(u)]_{u=a+z}^{u=b+z} d z,\ \sigma(z) \equiv \operatorname{sgn}(\operatorname{Im} z),
\end{align}
where $f(z)$ and $g(z)$ with $z=x+iy$ are meromorphic functions for $a \leq x \leq b$ and satisfy
\begin{align}
\lim_{h\to 0} \int_{a \pm ih}^{b \pm ih} [g(z) \pm f(z)]  dz =0.
\end{align}
The residue part is
\begin{align}
R[f(z), g(z)]=\pi i\left[\sum_{k} \underset{z=z_{g, k}}{\operatorname{Res}} g(z)+\sum_{k} \sigma\left(z_{f, k}\right) \underset{z=z_{f, k}}{\operatorname{Res}} f(z)\right]. \label{eq:residue}
\end{align}
$z_{f, k}$ and $z_{g, k}$ stand for the poles of $f(z)$ and $g(z)$, respectively.
In the following, we assume $f(z)$ is regular in $a<x<b$, and then the second term of Eq.~\eqref{eq:residue} vanishes.

\subsection{Integer}
First, we consider the APF for $f(n)$, where $n \in \mathbb{Z}$ is the index of summation.
We set $g(z)=-i\cot(\pi z)f(z)$, where the residue of $g(z)$ at $z=n$ is $-i f(n) /\pi$.
Then, Eq.~(\ref{eq:GAPF}) is
\begin{align}
 & \sum_{n=\lceil a \rceil }^{\lfloor b \rfloor} f(n)-\int_a^b dx f(x) -\left(\frac{1}{2}f(a)+\frac{1}{2}f(b)\ \mathrm{if}\ a,b \in \mathbb{Z}\right) \nonumber\\
&=\frac{1}{2} \int_{-i \infty}^{+i \infty}[\sigma(z)-i\cot(\pi u)] f(u)|_{u=a+z}^{u=b+z} d z \nonumber\\
  &=\frac{1}{2} \int_{0}^{+i \infty}[+1-i\cot(\pi u)] f(u)|_{u=a+z}^{u=b+z} d z
  +\frac{1}{2} \int_{-i \infty}^{0}[-1-i\cot(\pi u)] f(u)|_{u=a+z}^{u=b+z} d z, \label{eq:finite_APF}
\end{align}
where $\lceil x \rceil$ and $\lfloor x \rfloor$ are the ceiling function and the floor function, respectively.
If $z=a$ and/or $z=b$ are poles of $g(z)$, we have to avoid the poles on the integral path.
When we avoid the poles along a small semicircle, we obtain $-\frac{1}{2}f(a)$ and/or $-\frac{1}{2}f(b)$.
By using the exponential form of $\pm 1 -i \cot(\pi u)$,
\begin{align}
  \pm 1 -i \cot(\pi u)&=\pm\frac{ e^{ i \pi u}-e^{ -i \pi u}}{e^{i \pi u}-e^{-i \pi u}}+\frac{ e^{ i \pi u}+e^{ -i \pi u}}{e^{i \pi u}-e^{-i \pi u}} \nonumber\\
  &=\frac{ 2 e^{\pm i\pi u}}{e^{i \pi u}-e^{-i \pi u}} \nonumber\\
  &=\frac{\mp 2 }{e^{\mp 2i \pi u}-1},
\end{align}
the right-hand side of Eq.~(\ref{eq:finite_APF}) is written as
\begin{align}
  &+\frac{1}{2} \int_{0}^{+i \infty}dz \left[\frac{- 2f(u) }{e^{- 2i \pi u}-1}\right]_{u=a+z}^{u=b+z}
  +\frac{1}{2} \int_{-i \infty}^{0} dz \left[\frac{ 2f(u) }{e^{ 2i \pi u}-1} \right]_{u=b+z}^{u=a+z} \nonumber\\
  &=+i \int_{0}^{ \infty}dy \left[\frac{- f(u) }{e^{- 2i \pi u}-1}\right]_{u=a+iy}^{u=b+iy}
  -i\int_{ \infty}^{0}dy\left[\frac{ f(u) }{e^{ 2i \pi u}-1} \right]_{u=a-iy}^{u=b-iy} \nonumber\\
  &=i \int_{0}^{ \infty}dy \frac{ f(a+iy) }{e^{ 2 \pi (y-ia)}-1}
  -i \int_{0}^{ \infty}dy \frac{ f(a-iy) }{e^{ 2 \pi (y+ia)}-1}
  -i \int_{0}^{ \infty}dy \frac{ f(b+iy) }{e^{ 2 \pi (y-ib)}-1}
  +i \int_{0}^{ \infty}dy \frac{ f(b-iy) }{e^{ 2 \pi (y+ib)}-1}.
\end{align}
Finally, we obtain the APF for $f(n)$:
\begin{align}
  &\quad\sum_{n=\lceil a \rceil }^{\lfloor b \rfloor} f(n)-\int_a^b dx f(x)-\left(\frac{1}{2}f(a)+\frac{1}{2}f(b)\ \mathrm{if}\ a,b \in \mathbb{Z}\right) \nonumber\\
  &=i \int_{0}^{ \infty}dy \frac{ f(a+iy) }{e^{ 2 \pi (y-ia)}-1}
  -i \int_{0}^{ \infty}dy \frac{ f(a-iy) }{e^{ 2 \pi (y+ia)}-1}
  -i \int_{0}^{ \infty}dy \frac{ f(b+iy) }{e^{ 2 \pi (y-ib)}-1}
  +i \int_{0}^{ \infty}dy \frac{ f(b-iy) }{e^{ 2 \pi (y+ib)}-1}. \label{eq:FAPF_int}
\end{align}

\subsection{Half-integer}
Next, we consider the APF for $f(n+1/2)$, where $n+1/2$ ($n \in \mathbb{Z}$).
We set $g(z)=i\tan(\pi z)f(z)$, where the residue of $g(z)$ at $z=n+1/2$ is $-if(n+1/2)/\pi$.
Then, Eq.~(\ref{eq:GAPF}) is
\begin{align}
  &\quad \sum_{n=\lceil a-1/2 \rceil }^{\lfloor b-1/2 \rfloor} f(n+1/2)-\int_a^b dx f(x)-\left(\frac{1}{2}f(a)+\frac{1}{2}f(b)\ \mathrm{if}\ a-\frac{1}{2},b-\frac{1}{2} \in \mathbb{Z}\right) \nonumber\\
  \label{eq:half_int_GAP}
  &=\frac{1}{2} \int_{0}^{+i \infty}[+1+i\tan(\pi u)] f(u)|_{u=a+z}^{u=b+z} d z
  +\frac{1}{2} \int_{-i \infty}^{0}[-1+i\tan(\pi u)] f(u)|_{u=a+z}^{u=b+z} d z.
\end{align}
By using the exponential forms of $\pm1+ i \tan(\pi u)$,
\begin{align}
  \pm 1 +i \tan(\pi u)&=\pm\frac{ e^{ i \pi u}+e^{ -i \pi u}}{e^{i \pi u}+e^{-i \pi u}}+\frac{ e^{ i \pi u}-e^{ -i \pi u}}{e^{i \pi u}+e^{-i \pi u}} \nonumber\\
  &=\frac{\pm2 e^{ \pm i \pi u}}{e^{i \pi u}+e^{-i \pi u}} \nonumber\\
  &=\frac{\pm2 }{e^{\mp 2 i \pi u}+1},
\end{align}
the right-hand side of Eq.~(\ref{eq:half_int_GAP}) is
\begin{align}
  &\quad \frac{1}{2} \int_{0}^{+i \infty}[+1+i\tan(\pi u)] f(u)|_{u=a+z}^{u=b+z} d z
  +\frac{1}{2} \int_{-i \infty}^{0}[-1+i\tan(\pi u)] f(u)|_{u=a+z}^{u=b+z} d z \nonumber\\
  &=\int_{0}^{+i \infty}\left[\frac{f(u)}{e^{-2i\pi u}+1} \right]_{u=a+z}^{u=b+z} d z
  - \int_{-i \infty}^{0}
  \left[\frac{f(u)}{e^{+2i\pi u}+1} \right]_{u=a+z}^{u=b+z} d z \nonumber\\
  &=i\int_{0}^{\infty}\left[\frac{f(u)}{e^{-2i\pi u}+1} \right]_{u=a+iy}^{u=b+iy} d y-i\int^{ \infty}_{0}\left[\frac{f(u)}{e^{+2i\pi u}+1} \right]_{u=a-iy}^{u=b-iy} d y \nonumber\\
  &=-i\int_{0}^{\infty}\frac{f(a+iy)}{e^{2\pi (y-ia)}+1}  d y
  +i\int_{0}^{\infty}\frac{f(a-iy)}{e^{2\pi (y+ia)}+1}  d y
  +i \int_{0}^{\infty}\frac{f(b+iy)}{e^{2\pi (y-ib)}+1}  d y
  -i \int_{0}^{\infty}\frac{f(b-iy)}{e^{2\pi (y+ib)}+1} d y.
\end{align}
Finally, we obtain the APF for $f(n+1/2)$:
\begin{align}
  &\quad \sum_{n=\lceil a-1/2 \rceil }^{\lfloor b-1/2 \rfloor} f(n+1/2)-\int_a^b dx f(x)-\left(\frac{1}{2}f(a)+\frac{1}{2}f(b)\ \mathrm{if}\ a-\frac{1}{2},b-\frac{1}{2} \in \mathbb{Z} \right) \nonumber\\
 &=-i\int_{0}^{\infty}\frac{f(a+iy)}{e^{2\pi (y-ia)}+1} d y
 +i\int_{0}^{\infty}\frac{f(a-iy)}{e^{2\pi (y+ia)}+1} d y
 +i\int_{0}^{\infty}\frac{f(b+iy)}{e^{2\pi (y-ib)}+1} d y
 -i\int_{0}^{\infty}\frac{f(b-iy)}{e^{2\pi (y+ib)}+1} d y. \label{eq:FAPF_half}
\end{align}

\section{Derivation for naive lattice fermion} \label{App:3}
In this appendix, we derive the Casimir energy for massless naive lattice fermions from the APFs in finite range such as Eqs.~\eqref{eq:FAPF_int} and \eqref{eq:FAPF_half}.
From the Dirac operator~\eqref{eq:nf_D} at $am_f=0$, the dispersion relation in the $1+1$ dimensional space-time is
 \begin{align}
   a\sqrt{D_\mathrm{nf}^\dagger D_\mathrm{nf}}=\sqrt{\sin^2 ap_1(n)},
 \end{align}
where the momenta for the periodic and antiperiodic boundaries are $ap_1(n)=2\pi n/N$ and $ap_1(n)=2\pi (n+1/2)/N $, respectively.

After the analytic continuation ($n \to z$ for the periodic boundary and $n+1/2 \to z$ for the antiperiodic boundary), the resulting complex function $f(z=x+i y)=\sqrt{\sin^2\left(2\pi z/N \right)}$ has branch cuts along the $y$ direction at $x=0$, $x=N/2$, and $x=N$.
Therefore, to avoid the branch cut, we separately apply the APF to the two regions, $0 \leq x < N/2 $ and $N/2 < x < N$ (when the Brillouin zone is defined as $0 \leq ap_1 < 2\pi$).
For example, we consider a path along the branch cut at $x=N/2$, where the path is shifted by an infinitesimal parameter $\epsilon \rightarrow +0$ from the cut.
Then $f(z)$ on this path is (if $y>0$)
 \begin{align}
   \sin^2\left(\frac{2\pi}{N} \left(\frac{N}{2}+ \epsilon \pm iy\right)\right)& \simeq \left(\mp i\sinh \left(\frac{2\pi y}{N}\right)-\epsilon\right)^2 \nonumber\\
   &\simeq -\sinh^2\left(\frac{2\pi y}{N}\right)\pm i \epsilon, \\
   \sqrt{\sin^2\left(\frac{2\pi}{N} \left(\frac{N}{2}+ \epsilon \pm iy\right)\right)}& \simeq e^{\pm i \pi/2}\sinh\left(\frac{2\pi y}{N}\right)  \nonumber \\
   &=\pm i\sinh\left(\frac{2\pi y}{N}\right).
 \end{align}
Similarly, $f(z)$ on the paths at $x=N/2 -\epsilon$, $x=0+\epsilon$, and $x=N-\epsilon$ are
 \begin{align}
   \sqrt{\sin^2\left(\frac{2\pi}{N} \left(\frac{N}{2}- \epsilon \pm iy\right)\right)}&\simeq \mp i\sinh\left(\frac{2\pi y}{N}\right), \\
   \sqrt{\sin^2\left(\frac{2\pi}{N} \left(0+ \epsilon \pm iy\right)\right)}&\simeq \pm i\sinh\left(\frac{2\pi y}{N}\right), \\
   \sqrt{\sin^2\left(\frac{2\pi}{N} \left(N- \epsilon \pm iy\right)\right)}&\simeq \mp i\sinh\left(\frac{2\pi y}{N}\right).
 \end{align}
In the following, these expressions will be used for evaluating integrals.

\subsection{Periodic boundary}
For the periodic boundary, we substitute $f(z)=\sqrt{\sin^2(2\pi z/N)}$ into the APF for integers, Eq.~(\ref{eq:FAPF_int}).
We put $(a,b)=(0+\epsilon,N/2-\epsilon)$ in the first region and $(a,b)=(N/2+\epsilon,N-\epsilon)$ in the second region, and then
 \begin{align}
& aE_\mathrm{Cas}^\mathrm{1+1D,nf,P} \nonumber\\
   &= -i\int_0^\infty\frac{dy}{e^{2\pi y}-1}\left(\left.f(z)\right|^{z=\epsilon+iy}_{z=\epsilon-iy}-\left.f(z)\right|^{z=N-\epsilon+iy}_{z=N-\epsilon-iy}\right)
+i\int_0^\infty\frac{dy}{e^{ 2 \pi( y+i N/2)}-1}\left(\left.f(z)\right|^{z=N/2-\epsilon+iy}_{z=N/2-\epsilon-iy}-\left.f(z)\right|^{z=N/2+\epsilon+iy}_{z=N/2+\epsilon-iy}\right) \nonumber\\
   &=4\int_0^\infty\frac{dy\sinh(2\pi y/N)}{e^{2\pi y}-1}+4\int_0^\infty\frac{dy\sinh(2\pi y/N)}{e^{2\pi(y+iN/2)}-1}.
 \end{align}
For $e^{2\pi (y\pm i N/2)}$ of the denominator in the APF, we used $e^{+i N \pi}=e^{-i N \pi}$ for a integer $N$.
The contributions from the semicircles in the APF are zero because $f(0)=f(N/2)=f(N)=0$.
Here, the integrations are performed:
 \begin{align}
    \int_{0}^{ \infty}dy \frac{ \sinh(2\pi y /N) }{e^{ 2 \pi y}-1}
   &=+\frac{N}{4\pi}-\frac{1}{4}\cot\left(\frac{\pi}{N}\right), \label{eq:cot} \\
   \int_{0}^{ \infty}dy \frac{ \sinh(2\pi y /N) }{e^{ 2 \pi y}+1}
    &=-\frac{N}{4\pi}+\frac{1}{4}\csc\left(\frac{\pi}{N}\right). \label{eq:csc}
 \end{align}
Finally,
\begin{align}
   aE_\mathrm{Cas}^\mathrm{1+1D,nf,P} =
 \begin{cases}
    \frac{2N}{\pi} - \cot \frac{\pi}{2N} & (N = \mathrm{odd}) \\
    \frac{2N}{\pi} - 2\cot \frac{\pi}{N} & (N = \mathrm{even})
 \end{cases}.
\end{align}
For $N = \mathrm{odd}$, we used $\cot(x)+\csc(x)=\cot(x/2)$.

\subsection{Antiperiodic boundary}
For the antiperiodic boundary, we substitute $f(z)=\sqrt{\sin^2(2\pi z/N)}$ into the APF for half-integers, Eq.~(\ref{eq:FAPF_half}):
\begin{align}
aE_\mathrm{Cas}^\mathrm{1+1D,nf,AP}
=&i\int_0^\infty\frac{dy}{e^{2\pi y}+1}\left(\left.f(z)\right|^{z=\epsilon+iy}_{z=\epsilon-iy}-\left.f(z)\right|^{z=N-\epsilon+iy}_{z=N-\epsilon-iy}\right) \nonumber\\
&-i\int_0^\infty\frac{dy}{e^{ 2 \pi( y+i N/2)}+1}\left(\left.f(z)\right|^{z=N/2-\epsilon+iy}_{z=N/2-\epsilon-iy}-\left.f(z)\right|^{z=N/2+\epsilon+iy}_{z=N/2+\epsilon-iy}\right) \nonumber\\
  =&-4\int_0^\infty\frac{dy\sinh(2\pi y/N)}{e^{2\pi y}+1}-4\int_0^\infty\frac{dy\sinh(2\pi y/N)}{e^{2\pi(y+iN/2)}+1} \nonumber\\
  =& \begin{cases}
   \frac{2N}{\pi}-\cot \frac{\pi}{2N} & (N = \mathrm{odd})\\
   \frac{2N}{\pi}-2\csc \frac{\pi}{N} & (N = \mathrm{even})
   \end{cases}.
\end{align}
For the third equality, we used Eqs.~(\ref{eq:cot}) and (\ref{eq:csc}).

\section{Derivation for massless Wilson fermion} \label{App:4}
From the Dirac operator~\eqref{eq:Dw} at $r=1$ and $am_f=0$, the dispersion relations of massless Wilson fermion in the $1+1$-dimensional space-time is
\begin{align}
a \sqrt{D_\mathrm{W}^\dagger D_\mathrm{W}}= \sqrt{2-2\cos{ap_1(n)}} = 2\sqrt{\sin^2\left(\frac{ap_1(n)}{2}\right)},
\end{align}
where the momenta for the periodic and antiperiodic boundaries are $ap_1(n)=2\pi n/N$ and $ap_1(n)=2\pi (n+1/2)/N $, respectively.

In the dispersion relations after the analytic continuation, $f(z=x+i y)=2\sqrt{\sin^2\left(\pi z/N \right)}$ has a branch cut along the $y$ direction at $x=0$ and $x=N$.
$f(z)$ on paths along the branch cuts at $x=0$ and $x=N$ are (if $\epsilon \rightarrow +0,\ y>0$)
 \begin{align}
   \sin^2\left(\frac{\pi}{N}\left(0+\epsilon\pm i y\right)\right)
   &\simeq - \sinh^2\left(\frac{\pi y}{N}\right)\pm i\epsilon,\\
   \sqrt{\sin^2\left(\frac{\pi}{N} \left(0+ \epsilon \pm iy\right)\right)}&\simeq \pm i\sinh\left(\frac{\pi y}{N}\right),\\
   \sin^2\left(\frac{\pi}{N}\left(N-\epsilon\pm i y\right)\right)
   &\simeq - \sinh^2\left(\frac{\pi y}{N}\right)\mp i\epsilon,\\
   \sqrt{\sin^2\left(\frac{\pi}{N} \left(N- \epsilon \pm iy\right)\right)}&\simeq \mp i\sinh\left(\frac{\pi y}{N}\right).
 \end{align}

\subsection{Periodic boundary}
For the periodic boundary, we substitute $f(z)=2\sqrt{\sin^2(\pi z/N)}$ into the APF~(\ref{eq:FAPF_int}).
We put $(a,b)=(0+\epsilon,N-\epsilon)$, and then
 \begin{align}
aE_\mathrm{Cas}^{\mathrm{1+1D,W,P}} &= -i\int_0^\infty\frac{dy}{e^{2\pi y}-1}\left(\left.f(z)\right|^{z=\epsilon+iy}_{z=\epsilon-iy}-\left.f(z)\right|^{z=N-\epsilon+iy}_{z=N-\epsilon-iy}\right)\\
   &= 8\int_0^\infty\frac{dy\sinh(\pi y/N)}{e^{2\pi y}-1}= \frac{4N}{\pi} -2\cot \frac{\pi}{2N},
 \end{align}
where the contributions from the semicircles in the APF are zero because $f(0)=f(N)=0$.

\subsection{Antiperiodic boundary}
For the antiperiodic boundary, we substitute $f(z)=2\sqrt{\sin^2(\pi z/N)}$ into the APF~(\ref{eq:FAPF_half}),
\begin{align}
 aE_\mathrm{Cas}^{\mathrm{1+1D,W,AP}} &= i\int_0^\infty\frac{dy}{e^{2\pi y}+1}\left(\left.f(z)\right|^{z=\epsilon+iy}_{z=\epsilon-iy}-\left.f(z)\right|^{z=N-\epsilon+iy}_{z=N-\epsilon-iy}\right)\\
   &=-8\int_0^\infty\frac{dy\sinh(\pi y/N)}{e^{2\pi y}+1}= \frac{4N}{\pi} -2\csc \frac{\pi}{2N}.
\end{align}

\section{Derivation for Wilson fermion with negative mass $am_f=-2$} \label{App:5}
The dispersion relations of the Wilson fermion at $r=1$ and $am_f=-2$ in the $1+1$-dimensional space-time is
\begin{align}
  a\sqrt{D^\dagger D (am_f=-2)} = \sqrt{2+2\cos{ap_1(n)}}= 2\sqrt{\cos^2 \left( \frac{ap_1(n)}{2} \right)},
\end{align}
where the momenta for the periodic and antiperiodic boundaries are $ap_1(n)=2\pi n/N$ and $ap_1(n)=2\pi (n+1/2)/N $, respectively.

In the dispersion relations after the analytic continuation, $f(z=x+i y)=2\sqrt{\cos^2\left(\pi z/N \right)}$ has a branch cut along the $y$ direction at $x=N/2$.
Therefore, to avoid the branch cut, we separately apply the APF to the two regions, $0 \leq x < N/2 $ and $N/2 < x < N$ (when the Brillouin zone is defined as $0 \leq ap_1 < 2\pi$).
For $0 \leq x < N/2 $, the first and second terms in the right-hand side of the APF are zero because of $f(0+iy)-f(0-iy)=0$, and the third and fourth terms are nonzero.
For $N/2 < x < N$, the first and second terms are nonzero, and the third and fourth terms are zero because of $f(N+iy)-f(N-iy)=0$.
Note that these situations are different from the procedures shown for naive fermion and massless Wilson fermion since these fermions have the branch cuts at $x=0$ and $x=N$.

$f(z)$ on paths along the branch cut at $x=N/2$ are (if $\epsilon \rightarrow +0,\ y>0$)
\begin{align}
  \cos^2\left(\frac{\pi}{N}\left(\frac{N}{2}+\epsilon\pm i y\right)\right)
  &\simeq - \sinh^2\left(\frac{\pi y}{N}\right)\pm i\epsilon,\\
  \sqrt{\cos^2\left(\frac{\pi}{N}\left(\frac{N}{2}+\epsilon\pm i y\right)\right)}
  &\simeq \pm i \sinh\left(\frac{\pi y}{N}\right),\\
  \cos^2\left(\frac{\pi}{N}\left(\frac{N}{2}-\epsilon\pm i y\right)\right)
  &\simeq - \sinh^2\left(\frac{\pi y}{N}\right)\mp i\epsilon,\\
  \sqrt{\cos^2\left(\frac{\pi}{N}\left(\frac{N}{2}-\epsilon\pm i y\right)\right)}
  &\simeq \mp i \sinh\left(\frac{\pi y}{N}\right).
\end{align}

\subsection{Periodic boundary}
For the periodic boundary, we substitute $f(z)=2 \sqrt{\cos^2(\pi z/N)}$ into the APF for integers, Eq.~(\ref{eq:FAPF_int}).
We put $(a,b)=(0+\epsilon,N/2-\epsilon)$ in the first region and $(a,b)=(N/2+\epsilon,N-\epsilon)$ in the second region, and then
\begin{align}
  aE_\mathrm{Cas}^\mathrm{1+1D,P} (am_f=-2) &= -i\int_0^\infty\frac{dy}{e^{2\pi (y+iN/2)}-1}\left(\left.f(z)\right|^{z=N/2+\epsilon+iy}_{z=N/2+\epsilon-iy}-\left.f(z)\right|^{z=N/2-\epsilon+iy}_{z=N/2-\epsilon-iy}\right)\\
  &=8\int_0^\infty\frac{dy\sinh\left(\frac{\pi y}{N}\right)}{e^{2\pi (y+iN/2)}-1}=
  \begin{cases}
   \frac{4N}{\pi}-2\csc \frac{\pi}{2N} & (N = \mathrm{odd}) \\
   \frac{4N}{\pi}-2\cot \frac{\pi}{2N} & (N = \mathrm{even})
 \end{cases}.
\end{align}
The contributions from the semicircles around $z=0$ and $z=N$, namely, the third term in the left-hand side of the APF~(\ref{eq:FAPF_int}), is $\frac{1}{2}f(0)+\frac{1}{2}f(N)$.
The Casimir energy is defined as $\sum_{n=0}^{N-1} - \int_{0}^{N} dx f(x)$ where the Brillouin zone is defined as $0 \leq ap < 2\pi$, while the APF is now defined as $\sum_{n=0}^{N} f(n) -  \int_{0}^{N} dx f(x)$.
Therefore we have to subtract $f(N)$ from the APF in order to obtain the Casimir energy.
After we subtract $f(N)$ from $\frac{1}{2}f(0)+\frac{1}{2}f(N)$, we get $\frac{1}{2}f(0)-\frac{1}{2}f(N) = 0$.
On the other hand, the contribution from the semicircles around $z=N/2\ (N = \text{even}) $ disappears due to $f(N/2)=0$.

\subsection{Antiperiodic boundary}
For the antiperiodic boundary, we substitute $f(z)=2 \sqrt{\cos^2(\pi z/N)}$ into the APF for half-integers, Eq.~(\ref{eq:FAPF_half}),
\begin{align}
aE_\mathrm{Cas}^\mathrm{1+1D,AP} (am_f=-2)
&= i\int_0^\infty\frac{dy}{e^{2\pi (y+iN/2)}+1}\left(\left.f(z)\right|^{z=N/2+\epsilon+iy}_{z=N/2+\epsilon-iy}-\left.f(z)\right|^{z=N/2-\epsilon+iy}_{z=N/2-\epsilon-iy}\right)\\
  &=-8\int_0^\infty\frac{dy\sinh\left(\frac{\pi y}{N}\right)}{e^{2\pi (y+iN/2)}+1}=
  \begin{cases}
     \frac{4N}{\pi}-2\cot \frac{\pi}{2N}  & (N = \mathrm{odd}) \\
     \frac{4N}{\pi}-2\csc \frac{\pi}{2N}  & (N = \mathrm{even})
 \end{cases}.
\end{align}
The contributions from the semicircles around $z=N/2 \ ( N = \text{odd})$ are zero because $f(z=N/2)=0$.
\end{widetext}

\bibliography{casimir_ref}
\end{document}